%% file: ms.tex
\documentclass[12pt,modern,twocolumn,tighten]{aastex63}
\usepackage{amsmath,amstext,amssymb}
\usepackage[T1]{fontenc}
\usepackage{apjfonts}
\usepackage[figure,figure*]{hypcap}
\usepackage{graphics,graphicx}
\usepackage{hyperref}
\usepackage{natbib}
\usepackage[caption=false]{subfig} 
\usepackage{enumitem} 
\usepackage{epigraph}

\newcommand{\cn}{$\delta$\ Lyr\ cluster} 
\newcommand{\sn}{Kepler\,1627} 
\newcommand{\pn}{Kepler\,1627Ab} 

\newcommand{\clusterage}{$38^{+6}_{-5}$\,Myr} %

\newcommand{\nkinematic}{1{,}201} 
\newcommand{\noriginal}{3{,}071} 
\newcommand{\nwithtess}{924} 
\newcommand{\nkinwithtess}{391} 
\newcommand{\nkinwithtessandcrowding}{192} 
\newcommand{\nkindefaultcleaning}{145} 
\newcommand{\nrotgood}{135} 
\newcommand{\nfracprot}{70} 

\newcommand{\kms}{\,km\,s$^{-1}$}
\newcommand{\ms}{\,m\,s$^{-1}$}
\newcommand{\bpmrpo}{(G_{\rm BP}-G_{\rm RP})_0}

\received{2021 Oct 12}
\revised{2021 Dec 6}
\accepted{2021 Dec 27}
\shorttitle{Kepler\,1627}

\begin{document}

\title{
  A 38 Million Year Old Neptune-Sized Planet in the Kepler Field
}

\input{authors.tex}

\begin{abstract}
  Kepler\,1627A is a G8V star previously known to host a
  $3.8\,R_\oplus$ \replaced{mini-Neptune}{planet} on a 7.2\,day orbit.  The star was
  observed by the Kepler space telescope because it is nearby
  ($d=329\,{\rm pc}$) and it resembles the Sun.  Here we show using
  Gaia kinematics, TESS stellar rotation periods, and spectroscopic
  lithium abundances that Kepler\,1627 is a member of the \clusterage\
  old $\delta$~Lyr cluster.  To our knowledge, this makes
  Kepler\,1627Ab the youngest planet with a precise age yet found by
  the \replaced{main}{prime} Kepler mission.  The Kepler photometry shows two
  peculiarities: the average transit profile is asymmetric, and the
  individual transit times might be correlated with the local light curve
  slope.  We discuss possible explanations for each anomaly.  More
  importantly, the $\delta$~Lyr cluster
  is one of $\sim$10$^3$ coeval groups whose properties have been clarified
  by Gaia.  Many other exoplanet hosts are candidate members
  of these clusters; their ages can be verified \replaced{through}{with} the
  trifecta of Gaia, TESS, and ground-based spectroscopy.
\end{abstract}

\keywords{
  exoplanet evolution (491),
  open star clusters (1160),
	stellar ages (1581)
}


\section{Introduction}

While thousands of exoplanets have been discovered orbiting nearby
stars, the vast majority of them are several billion years old.  This
makes it difficult to test origin theories for the different families
of planets, since many evolutionary processes are expected to operate
on timescales of less than 100 million years.

For instance, the ``mini-Neptunes'', thought to be made of metal
cores, silicate mantles \citep{kite_atmosphere_2020}, and extended
hydrogen-dominated atmospheres, are expected to shrink
in size by factors of several over their first $10^8$ years.
Specifically, in the models of \citet{owen_atmospheres_2016} and
\citet{owen_constraining_2020}, the $\approx$$5\,M_\oplus$ planets start
with sizes of 4\,--\,12\,$R_\oplus$ shortly after the time of disk
dispersal ($\lesssim$$10^7$\,years), and shrink to sizes of
2\,--\,4\,$R_\oplus$ by 10$^8$ years.  While the majority of this change
is expected to occur within the first few million years after the disk
disperses \citep{ikoma_situ_2012}, stellar
irradiation and internal heat can also power gradual outflows
which, if strong enough, can deplete or entirely strip the atmosphere \citep{lopez_how_2012,Owen_Wu_2013,ginzburg_corepowered_2018}.\added{ The photoevaporative and core-powered
	outflows are thought to persist for $\approx$$10^8$ to $\approx$$10^9$ years,
	though the details depend on the planetary masses,
	the irradiation environments, and the initial atmospheric
	mass fractions \citep{owen_evaporation_2017,gupta_signatures_2020,2021MNRAS.503.1526R,2021MNRAS.501L..28K}.} Discovering young planets, measuring their masses, and detecting their
atmospheric outflows are key steps toward testing this paradigm, which
is often invoked to explain the observed radius distribution of mature
exoplanets \citep{Fulton_et_al_2017,van_eylen_asteroseismic_2018}.

The K2 and TESS missions have now enabled the detection of about ten
close-in planets younger than 100 million years, all smaller than
Jupiter
\citep{Mann_K2_33b_2016,David_et_al_2017,david_four_2019,newton_tess_2019,bouma_cluster_2020,plavchan_planet_2020,rizzuto_tess_2020,martioli_aumicbc_2021}.
The Kepler mission however has not yielded any planets with precise
ages below one gigayear \citep{Meibom_et_al_2013}.  The reason is that
during the \replaced{main}{prime} Kepler mission (2009--2013), only four open clusters
were known in the Kepler field,
with ages spanning 0.7\,Gyr to 9\,Gyr \citep{meibom_kepler_2011}.
Though isochronal, gyrochronal, and lithium-based analyses suggest
that younger Kepler planets do exist
\citep{walkowicz_rotation_2013,berger_identifying_2018,david_sizes_2021}, accurate and precise
age measurements typically require an ensemble of stars.  Fortunately,
recent analyses of the Gaia data have greatly expanded our knowledge
of cluster memberships \citep[{\it
e.g.},][]{CantatGaudin2018a,Zari2018,KounkelCovey2019,Meingast2021,Kerr2021}.
As part of our Cluster Difference Imaging Photometric Survey (CDIPS,
\citealt{bouma_cdipsI_2019}), we concatenated the available analyses
from the literature, which yielded a list of candidate young and
age-dated stars (see Appendix~\ref{app:targetlist}).

Matching our young star list against stars observed by Kepler revealed
that Kepler observed a portion of the \cn\ (Stephenson-1; Theia~73).
More specifically, a clustering analysis of the Gaia data by
\citet{KounkelCovey2019} reported that Kepler~1627 (KIC 6184894; KOI
5245) is a \cn\ member.  Given the previous statistical
validation of the close-in \replaced{mini-Neptune}{Neptune-sized
planet} Kepler 1627b
\citep{2012ApJS..199...24T,morton_false_2016,thompson_planetary_2018},
we begin by scrutinizing the properties of the cluster
(Section~\ref{sec:cluster}).  We find that the $\delta$ Lyr cluster is
\clusterage\ old, and in Section~\ref{sec:stars} show that \sn\ is
both a binary and also a member of the cluster.  Focusing on the
planet (Section~\ref{sec:planet}), we confirm that despite the
existence of the previously unreported \replaced{M2.5}{M3}V companion, hereafter
Kepler\,1627B, the planet orbits the G-dwarf primary, Kepler\,1627A.
We also analyze an asymmetry in the average transit profile, and a
possible correlation between the individual transit times and the
local light curve slope.  We conclude by discussing broader
implications for our ability to age-date a larger sample of planets
(Section~\ref{sec:conc}).

\section{The Cluster}
\label{sec:cluster}

\begin{figure*}[t]
	\begin{center}
		\leavevmode
		\includegraphics[width=0.99\textwidth]{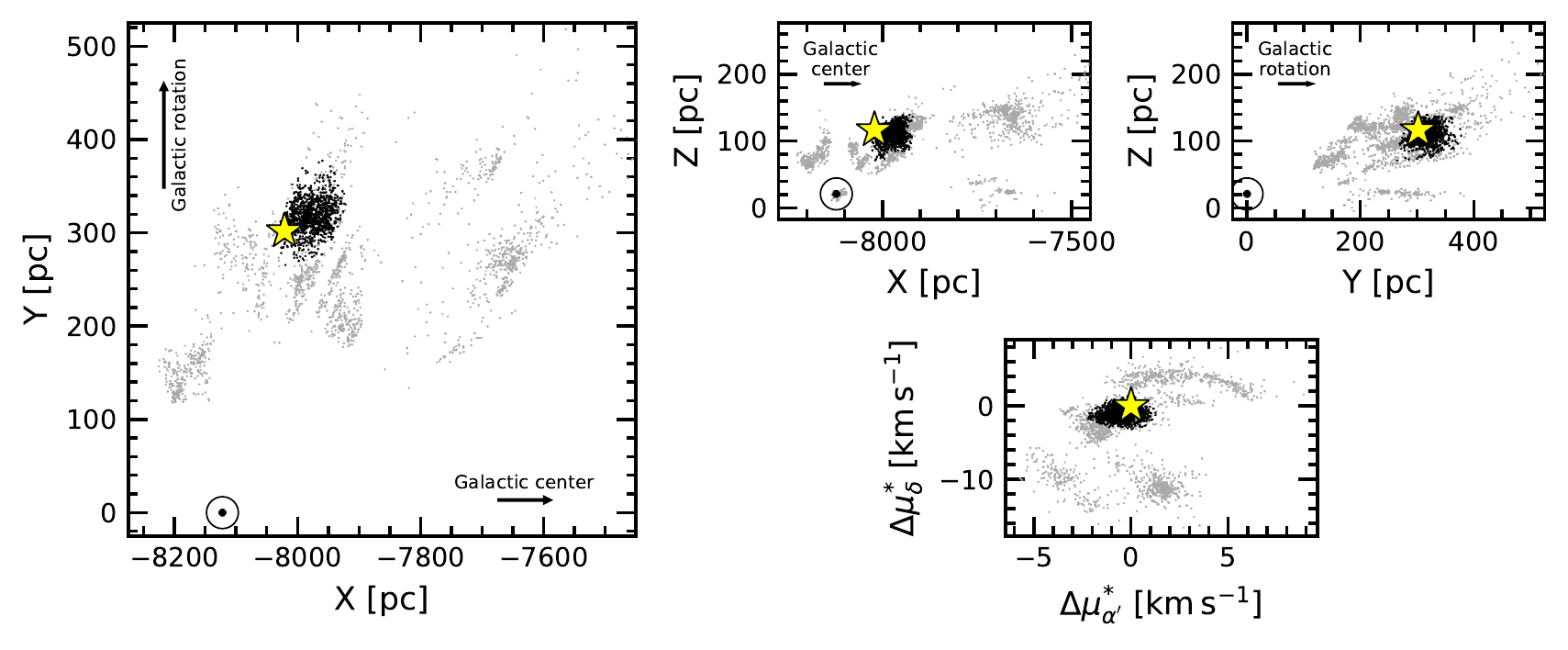}
	\end{center}
	\vspace{-0.7cm}
	\caption{
    {\bf Galactic positions  and tangential velocities of stars in the
    $\delta$\,Lyr cluster.} Points are reported cluster members from
    \citet{KounkelCovey2019}.  The tangential velocities relative to
    \sn\ (bottom right) are computed assuming that every star has the
    same three-dimensional spatial velocity as \sn.  Our analysis
    considers stars (black points) in the spatial and kinematic
    vicinity of Kepler\,1627 (yellow star).  The question of whether
    the other candidate cluster members (gray points) are part of the
    cluster is outside our scope.  The location of the Sun is
    ($\odot$) is shown \added{to clarify the direction along which
    parallax uncertainties are expected to produce erroneous
    clusters}.
		\label{fig:XYZvtang}
	}
\end{figure*}

To measure the age of the $\delta$ Lyr cluster, we first selected a
set of candidate cluster members
(Section~\ref{sec:kinematicselection}), and then analyzed these stars
using a combination of the isochronal and gyrochronal techniques
(Section~\ref{sec:clusterage}).

\subsection{Selecting Cluster Members}
\label{sec:kinematicselection}

\citet{KounkelCovey2019} applied an unsupervised clustering algorithm
to Gaia DR2 on-sky positions, proper motions, and parallaxes for stars
within the nearest kiloparsec.  For the \cn\ (Theia 73), they reported
\noriginal\ candidate members.  We matched these stars against the
latest Gaia EDR3 observations using the \texttt{dr2\_neighbourhood}
table from the ESA archive,
taking the stars closest in proper motion and epoch-corrected angular
distance as the presumed match \citep{gaia_collaboration_2021_edr3}.
\replaced{In Figure~\ref{fig:XYZvtang}, \added{we }have shown galactic positions only for
the stars with parallax signal-to-noise exceeding 20. }{For plotting purposes, we focused only
on the stars with parallax signal-to-noise exceeding 20.} \added{We
calculated the tangential velocities for each of these stars relative to \sn\
($\Delta \mu^{*}$) by subtracting the observed proper motion from what the proper motion at each star's
position would be if it were co-moving with \sn. }

\begin{figure}[t]
	\begin{center}
		\leavevmode
		\includegraphics[width=0.47\textwidth]{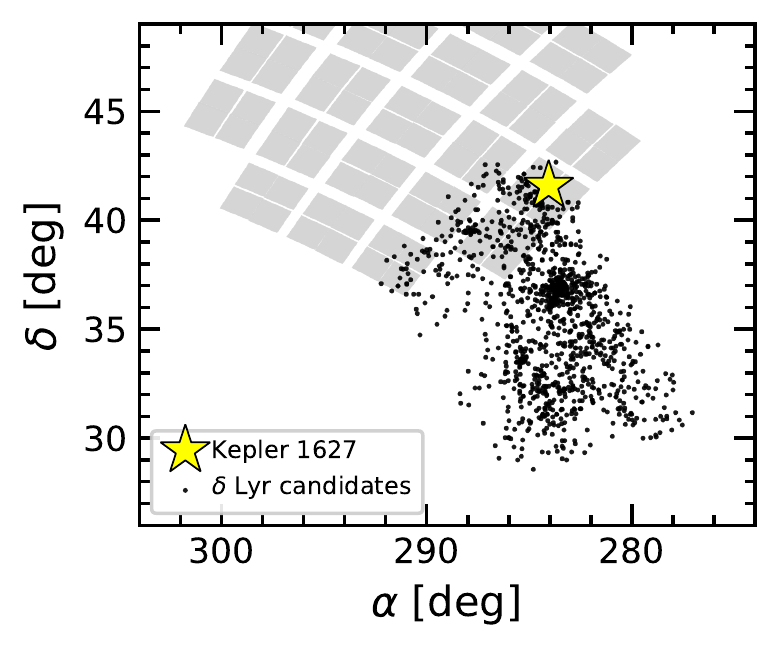}
	\end{center}
	\vspace{-0.7cm}
	\caption{
    {\bf Kepler's view of the $\delta$~Lyr cluster, shown in equatorial
    	coordinates.} Each black circle
    is a candidate cluster member selected based on its position and kinematics (Figure~\ref{fig:XYZvtang}).  Of
    the \nkinematic\ candidate cluster members, 58 have at least one
    quarter of Kepler data.  TESS has also observed most of the
    cluster, for one to two lunar months to date.
		\label{fig:skychart}
	}
\end{figure}

\replaced{The}{Figure~\ref{fig:XYZvtang} shows that the} reported
cluster members (gray and black points) extend over a much larger
volume \added{in both physical and kinematic space }than the cluster previously identified by
\citet{stephenson_possible_1959} and later corroborated by
\citet{eggen_photometric_1968}.  While the non-uniform ``clumps'' of
stars might comprise a {\it bona fide} cluster of identically-aged
stars, they could also be heavily contaminated by field stars.
\added{One reason to suspect this is that the spread in tangential
velocities exceeds typical limits for kinematic coherence
\citep{Meingast2021}}.
We
therefore considered stars only in the immediate kinematic and spatial
vicinity of \sn\ as candidate cluster members.  We performed
\replaced{the}{this}
selection cut\deleted{s} manually, by drawing lassos with the interactive
\texttt{glue} visualization tool \citep{beaumont_2014_13866} in the
four projections shown in Figure~\ref{fig:XYZvtang}.  The overlap
between the Kepler field and the resulting candidate cluster members
is shown in Figure~\ref{fig:skychart}.  While this method will include
some field interlopers in the ``cluster star'' sample, and vice-versa,
it should suffice for our aim of verifying the existence of the
cluster in the vicinity of \sn.

\subsection{The Cluster's Age}
\label{sec:clusterage}

\subsubsection{Color-Absolute Magnitude Diagram}
\label{sec:camd}

\begin{figure*}[tp]
	\begin{center}
		\leavevmode
		\subfloat{
			\includegraphics[width=0.49\textwidth]{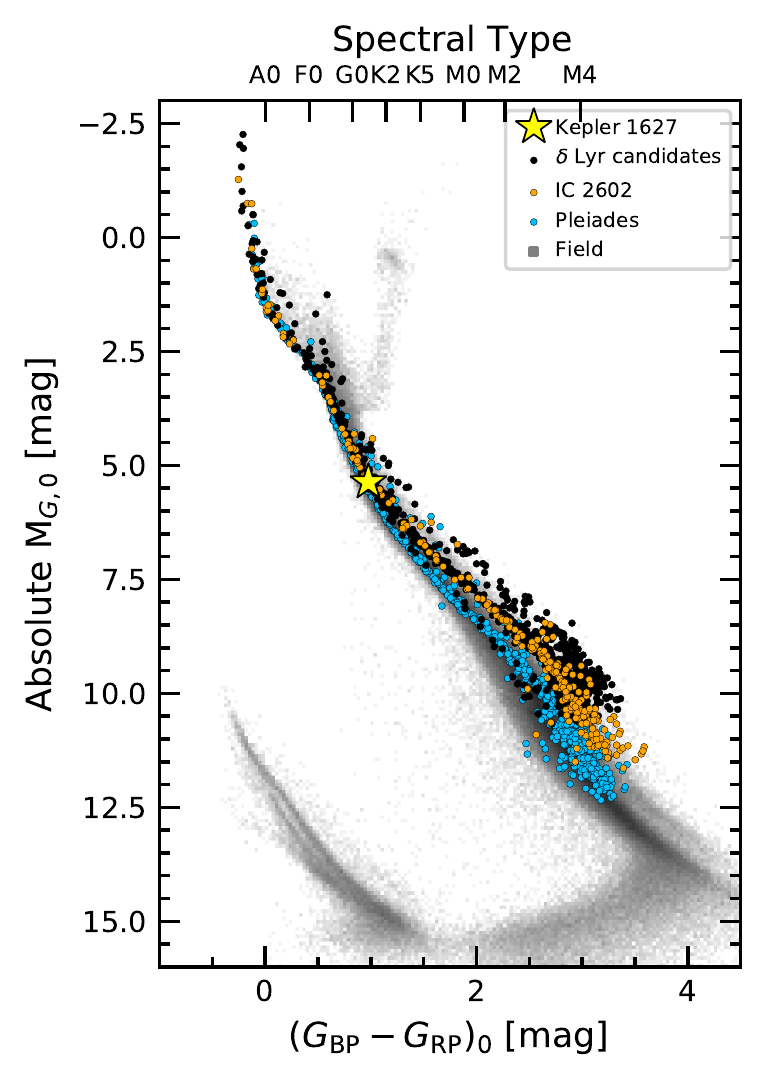}
			\includegraphics[width=0.469\textwidth]{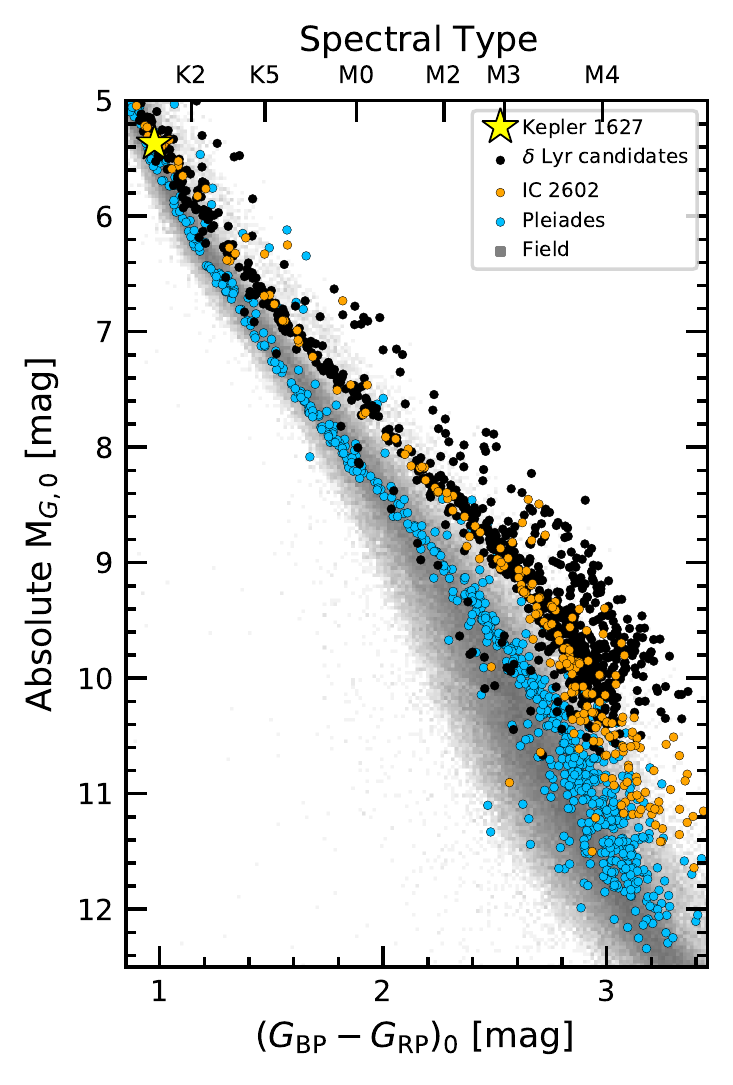}
		}
		
		\vspace{-0.6cm}
		\subfloat{
			\includegraphics[width=0.7\textwidth]{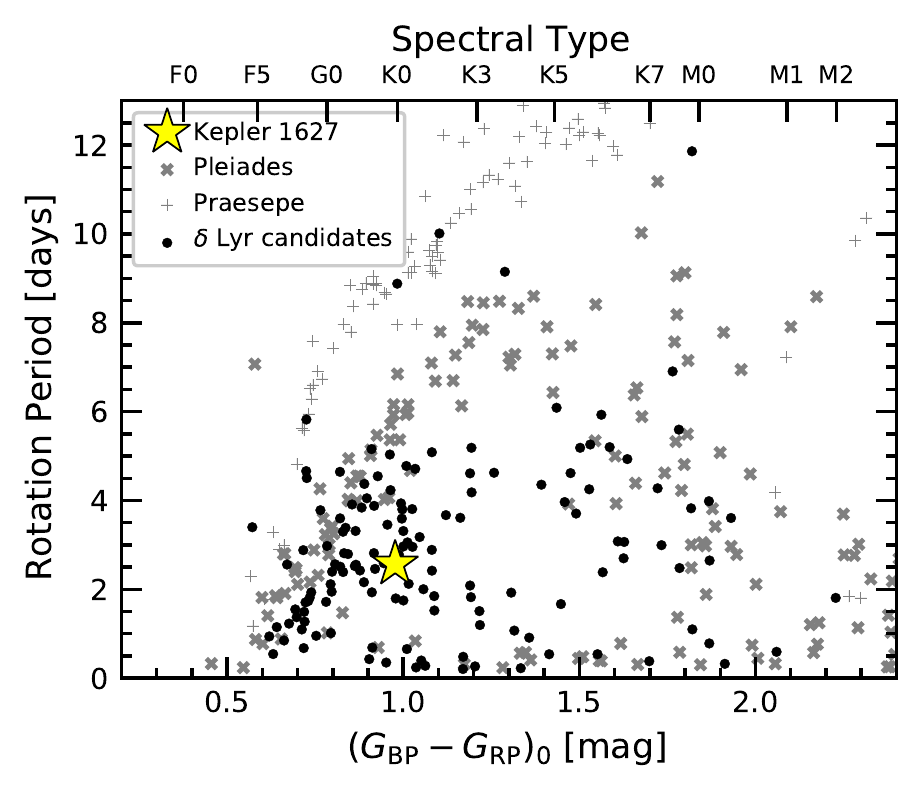}
		}
	\end{center}
	\vspace{-0.7cm}
	\caption{
		{\bf The \cn\ is \clusterage\ old.}  {\it Top:} Color-absolute
		magnitude diagram of candidate \cn\ members, in addition to stars
		in IC\,2602 ($\approx38$\,Myr), the Pleiades ($\approx 115$\,Myr),
		and the Gaia EDR3 Catalog of Nearby Stars (gray background).  The
		zoomed right panel highlights the pre-main-sequence.  The \cn\ and
		IC\,2602 are approximately the same isochronal age.  {\it Bottom:}
		TESS and Kepler stellar rotation period versus dereddened Gaia
		color, with the Pleiades and Praesepe (650\,Myr) shown for
		reference \citep{rebull_rotation_2016a,douglas_poking_2017}.  Most
		candidate \cn\ members are gyrochronally younger than the
		Pleiades; outliers are probably field interlopers.
		\label{fig:age}
	}
\end{figure*}

We measured the isochrone age using an empirical approach.  The \added{upper }left
panel of Figure~\ref{fig:age} shows the color-absolute magnitude
diagram (CAMD) of candidate \cn\ members, IC\,2602\added{ ($\approx 38$\,Myr)}, the Pleiades\added{ ($\approx 115$\,Myr)}, and
the field. The stars from the Pleiades and IC\,2602 were adopted from
\citet{CantatGaudin2018a}, and the field stars are from the Gaia EDR3
Catalog of Nearby Stars \citep{gaia_gcns_2021}.\added{ We also
compared against $\mu$-Tau ($62\pm7$\,Myr; \citealt{gagne_mutau_2020})
and the Upper-Centaurus-Lupus (UCL) component of the Sco OB2
association ($\approx$$16$\,Myr; \citealt{pecaut_star_2016}).
We adopted the UCL members from \citet{Damiani2019}.
For visual clarity, the latter two clusters are not shown in Figure~\ref{fig:age}.} We cleaned \replaced{these}{the membership lists}
following the data filtering criteria from \citet[][Appendix
B]{GaiaCollaboration2018}, except that we weakened the parallax
precision requirement to $\varpi/\sigma_\varpi>5$.  \added{This also involved cuts
on the photometric signal to noise ratio, the number of visibility
periods used, the astrometric $\chi^2$ of the single-source solution, and the
$G_{\rm BP}-G_{\rm RP}$ color excess factor.} These filters were
designed to include genuine binaries while omitting instrumental
artifacts.  

\replaced{We then corrected for extinction by querying}{To correct for extinction, we queried} the
3-dimensional maps of \citet{capitanio_threedimensional_2017} and
\citet{lallement_threedimensional_2018}\footnote{\url{https://stilism.obspm.fr/},
2021/09/25}, and applied the extinction coefficients $k_X\equiv
A_X/A_0$ computed by \citet{GaiaCollaboration2018} assuming that $A_0
= 3.1 E(B-V)$.  For \added{UCL, }IC\,2602, the Pleiades, and the \cn, this
procedure yielded a respective mean and standard deviation for the
reddening of \replaced{$E(B-V)=\{0.020\pm0.003, 0.045\pm0.008,
0.032\pm0.006\}$}{$E(B-V)=\{0.084\pm0.041, 0.020\pm0.003, 0.045\pm0.008,
0.032\pm0.006\}$}.
These values \replaced{agree reasonably well with}{are within a
factor of two of} previously reported values
\replaced{from}{in} the literature \citep{pecaut_star_2016,GaiaCollaboration2018,KounkelCovey2019,bossini_age_2019}\added{, and are all small enough that the choice of whether to use them {\it vs.} other
extinction estimates does not affect our primary conclusions}.

%
%
%

Figure~\ref{fig:age} shows that the \cn\ and IC\,2602 overlap, and
therefore are approximately the same age.  \replaced{In our exploration, we also
compared against $\mu$-Tau ($62\pm7$\,Myr; \citealt{gagne_mutau_2020})
and the Upper-Centaurus-Lupus (UCL) component of the Sco OB2
association ($\approx$$16$\,Myr; \citealt{pecaut_star_2016}). The
pre-main-sequence M dwarfs of the \cn\ were intermediate between the
latter two clusters.}{The
pre-main-sequence M dwarfs of a fixed color in the \cn\ are more luminous than those of the Pleiades, and were also seen to be 
less luminous than those in
UCL.}  To turn this heuristic interpolation
into \replaced{an age}{a quantitative age}
measurement, we used the empirical method developed by
\citet{gagne_mutau_2020}.  In brief, we fitted the pre-main-sequence
loci of a set of reference clusters, and the locus of the target \cn\
was then modeled as a piecewise linear combination of these reference
clusters.  For our reference clusters, \replaced{we adopted members of UCL,
IC\,2602, and the Pleiades from \citet{Damiani2019} and
\cite{CantatGaudin2018a} respectively}{we used UCL, IC\,2602, and the
Pleiades}.  We removed binaries by
requiring $\texttt{RUWE}<1.3$, $\texttt{radial\_velocity\_error}$
below the 80$^{\rm th}$ percentile of each cluster's distribution, and
excluded stars that were obvious photometric binaries in the CAMD.\footnote{
  For a description of the renormalized unit weight error (RUWE), see the GAIA DPAC technical note
  \url{http://www.rssd.esa.int/doc_fetch.php?id=3757412}.}  We then passed a
moving box average and standard deviation across the CAMD in 0.10\,mag bins,
fitted a univariate spline to the binned values, and assembled a
piecewise grid
of \deleted{hybrid }isochrones spanning the ages between UCL \replaced{to}{and} the Pleiades using
Equation\replaced{s~6 and~7}{~6} from
\citet{gagne_mutau_2020}.  
\added{To derive a probability distribution function for the age of \cn, we
then assumed a Gaussian likelihood that treated the interpolated isochrones as
the ``model'' and the $\delta$\ Lyr\ cluster's isochrone as the ``data''
(Equation~7 from \citealt{gagne_mutau_2020}).  The cluster's age and its
statistical uncertainty are then quoted as the the mean and standard deviation
of this age posterior.}


The ages returned by this procedure depend on the ages assumed for
each reference cluster.  We adopted a 115\,Myr age for the Pleiades
\citep{dahm_2015}, and a 16\,Myr age for UCL \citep{pecaut_star_2016}.
The age of IC\,2602 however is the most important ingredient, since it
receives the most weight in the interpolation.  Plausible ages for
IC\,2602 span 30\,Myr to 46\,Myr, with older ages being preferred by
the lithium-depletion-boundary (LDB) measurements
\citep{dobbie_ic_2010,randich_gaiaeso_2018} and younger ages by the
main-sequence turn-off
\citep{stauffer_rotational_1997,david_ages_2015,bossini_age_2019}.  If
we were to adopt the 30\,Myr age for IC\,2602, then the \cn\ would be
$31^{+5}_{-4}$\,Myr old.  For the converse extreme of 46\,Myr, the
\cn\ would be $44^{+8}_{-7}$\,Myr old.  We adopt an intermediate
38\,Myr age for IC\,2602, which yields an age for the \cn\ of
\clusterage.\footnote{Our exploration of the PARSEC and MIST isochrone
models over a grid of ages, metallicities, and reddenings, yielded the
best agreement for this $\approx$$38\,$Myr age as well, given ${\rm
[Fe/H] = +0.1}$ and $A_V=0.2$
\citep{bressan_parsec_2012,choi_mesa_2016};  this preferred CAMD
reddening is higher than the \citet{lallement_threedimensional_2018}
value by a factor of two.  } Follow-up studies of the LDB or
main-sequence turn-off in the \cn\ could help determine a more precise
and accurate age for the cluster, and are left for future work.

\subsubsection{Stellar Rotation Periods}

Of the \noriginal\ candidate \cn\ members reported by
\citet{KounkelCovey2019}, \nwithtess\ stars were amenable to rotation
period measurements ($G<17$ and $\bpmrpo>0.5$) using the TESS full
frame image data.\added{ As a matter of
scope, we restricted our attention to the \nkinwithtess\ stars
discussed in Section~\ref{sec:kinematicselection} in the spatial and
kinematic proximity of Kepler\,1627.} We extracted light curves from
the TESS images using the nearest pixel to each star, and regressed
them against systematics with the causal pixel model implemented in
the \texttt{unpopular} package \citep{hattorio_2021_cpm}.  We then
measured candidate rotation periods using a Lomb-Scargle periodogram
\citep{lomb_1976,scargle_studies_1982,astropy_2018}.  To enable cuts
on crowding, we queried the Gaia source catalog for stars within a
$21\farcs0$ radius of the target star (a radius of 1 TESS pixel).
Within this radius, we recorded the number of stars with greater
brightness than the target star, and with brightness within 1.25 TESS magnitudes of the target star.  

We then cleaned the candidate TESS rotation period measurements
through a combination of automated and manual steps.\deleted{ As a
matter of scope, we restricted our attention to the \nkinwithtess\
stars discussed in Section~\ref{sec:kinematicselection} in the spatial
and kinematic proximity of Kepler\,1627.}\deleted{ Kepler rotation
periods were derived by \citet{mcquillan_rotation_2014} for 28 of
our \nkinwithtess\ stars; for these cases, we
simply adopted the Kepler rotation period.}\added{ First, to validate
the TESS rotation periods, we compared against 28 stars
from \citet{mcquillan_rotation_2014} that were also observed by Kepler.  Of the 23 stars with Kepler
periods below 10 days, 21 of the TESS periods agreed with the Kepler
rotation periods; the other 2 were measured at the double-period
harmonic.  Of the remaining 5 stars with Kepler rotation periods above
10 days, none were correctly recovered by TESS, and 3 were near the
half-period harmonic.  We were therefore 
wary of any TESS rotation periods exceeding 10 days, and used the
Kepler rotation periods whenever possible.}
For the remaining stars with only TESS data, we
focused only on the stars for which no companions were known with a brightness
exceeding one-tenth of the target star in a $21\farcs0$ radius.  There
were \nkinwithtessandcrowding\ stars that met these crowding requirements, and
that had TESS data available.  For plotting purposes we then imposed a
selection based on the strength of the signal itself: we required the
Lomb Scargle power to exceed 0.2, and the period to be below 15\,days.

The lower panel of Figure~\ref{fig:age} shows the resulting
\nkindefaultcleaning\ stars.  
The majority of these stars fall below
the ``slow sequence'' of the Pleiades, consistent with a gyrochronal
age for the \cn\ below 100\,Myr.  In fact, the rotation-color
distributions of other 30\,Myr to 50\,Myr clusters ({\it e.g.},
IC\,2602 and IC\,2391) are indistinguishable
\citep{douglas_stephanie_t_2021_5131306}.  Approximately 10 of the
\cn\ stars appear as outliers above the ``slow sequence''.  Assuming
that they are all false positives ({\it i.e.}, field interlopers), our
rotation period detection fraction would be
$\nrotgood/\nkinwithtessandcrowding \approx \nfracprot\%$.   \replaced{The other
stars are likely to be field contaminants.}{Although some of these outlier stars might be
unresolved F+K binaries that are in the cluster
\citep{stauffer_rotation_2016}, assuming that they are field contaminants
provides a more secure lower bound of the rotation period detection fraction.}  A final possible
confounding factor -- binarity -- is known to affect the ``fast
sequence'' of stars beneath the slow sequence
\citep{meibom_effect_2007,gillen_ngts_2020,bouma_2021_ngc2516}.  We do
not expect it to change the central conclusion regarding the cluster's
age.

\section{The Stars}
\label{sec:stars}

\begin{figure*}[tp]
	\begin{center}
		\leavevmode
		\subfloat{
			\includegraphics[width=0.49\textwidth]{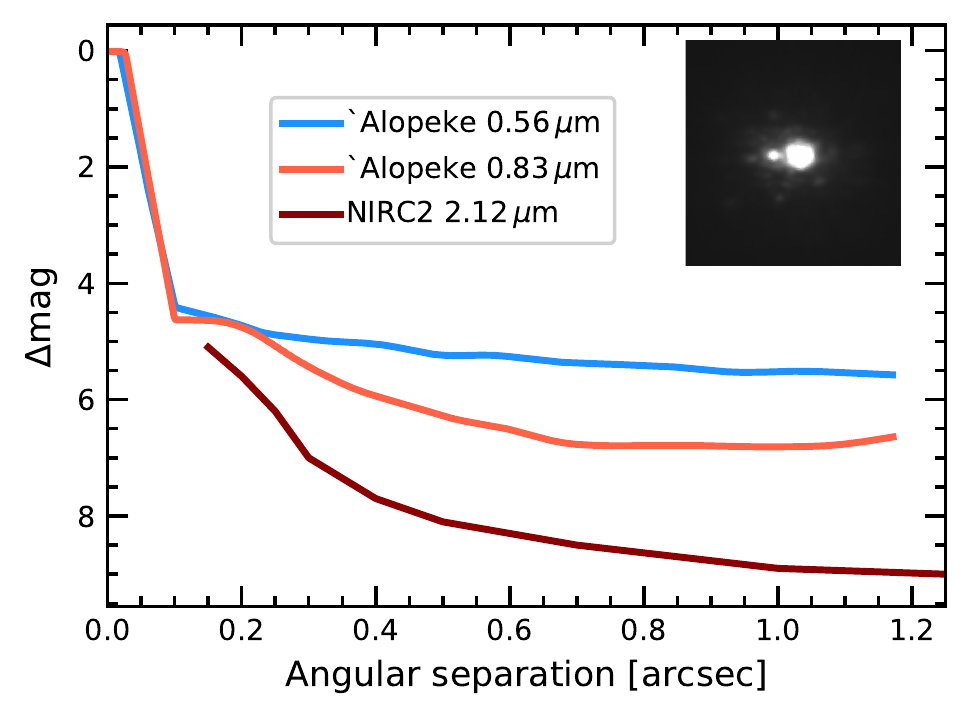}
			\includegraphics[width=0.48\textwidth]{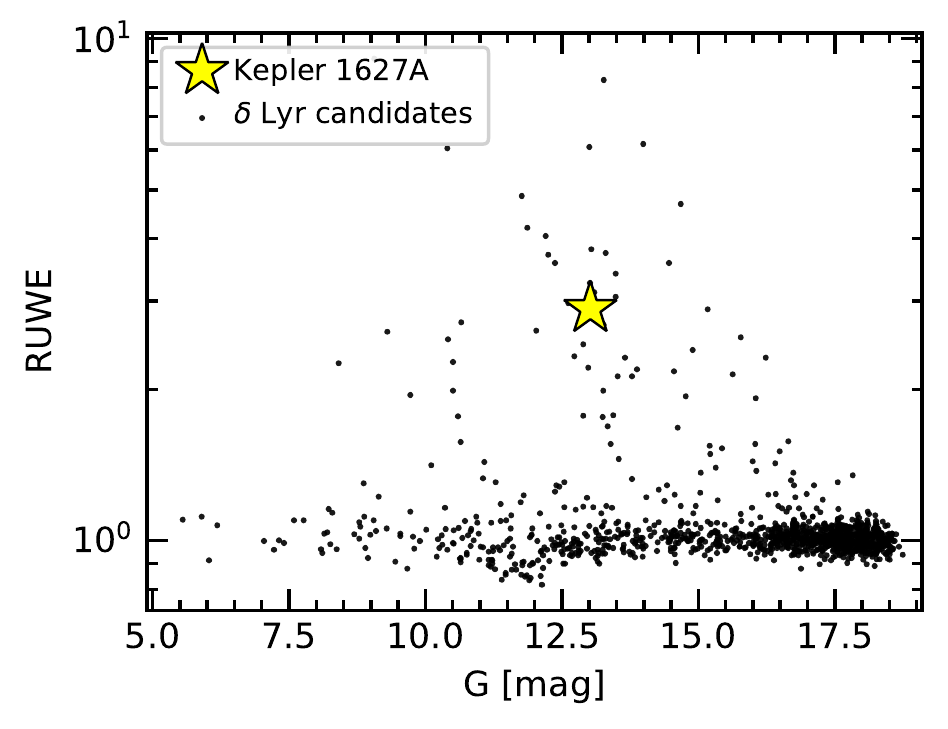}
		}
	\end{center}
	\vspace{-0.5cm}
	\caption{
		{\bf Kepler 1627 is a binary.} {\it Left:} High-resolution imaging
		from Gemini-North/`Alopeke and Keck/NIRC2 shows an $\approx$\replaced{M2.5}{M3}V
		companion at $\rho \approx 0\farcs16$, which corresponds to a
		projected separation of $53\pm4$\,AU.  The inset shows a cutout of
		the stacked NIRC2 image (North is up, East is left, scale is set
		by the separation of the binary).  The lines show 5-$\sigma$
		contrast limits for the `Alopeke filters, and 6-$\sigma$ contrast
		limits for NIRC2 outside of $0\farcs15$. {\it Right:} Gaia EDR3
		renormalized unit weight error (RUWE) point estimates for
		candidate $\delta$\,Lyr cluster members.  Since other members of
		the cluster with similar brightnesses have comparable degrees of
		photometric variability, the high RUWE independently suggests that
		Kepler\,1627 is a binary. 
		\label{fig:kep1627binary}
	}
\end{figure*}

\subsection{Kepler\,1627A}
\subsubsection{Age}
Based on the spatial and kinematic association of \sn\ with the \cn,
and the assumption that the planet formed shortly after the star, it
seems likely that \sn\ is the same age as the cluster. There are two
consistency checks on whether this is true: rotation and lithium.
Based on the Kepler light curve, the rotation period is
$2.642\pm0.042$\,days, where the quoted uncertainty is based on the
scatter in rotation periods measured from each individual Kepler
quarter.  This is consistent with comparable cluster members
(Figure~\ref{fig:age}).

To infer the amount of Li\,\textsc{I} from the 6708\,\AA\ doublet
\citep[{\it e.g.},][]{soderblom_ages_2014}, we acquired an iodine-free
spectrum from Keck/HIRES on the night of 2021 March 26 using the
standard setup and reduction techniques of the California Planet
Survey \citep{howard_cps_2010}.  Following the equivalent width
measurement procedure described by \citet{bouma_2021_ngc2516}, we find
${\rm EW}_{\rm Li} = 233^{+5}_{-7}$\,m\AA.   This value does not
correct for the Fe\,\textsc{I} blend at 6707.44\AA.  Nonetheless,
given the stellar effective temperature (Table~\ref{tab:starparams}),
this measurement is in agreement with expectations for a
$\approx40$\,Myr star ({\it e.g.}, as measured in IC~2602 by
\citealt{randich_gaiaeso_2018}).  It is also larger than any lithium
equivalent widths measured by \citet{berger_identifying_2018} in their
analysis of 1{,}301 Kepler-star spectra.

\subsubsection{Stellar Properties}

The adopted stellar parameters are listed in
Table~\ref{tab:starparams}.  The stellar mass, radius, and effective
temperature are found by interpolating against a 38$\,$Myr MIST
isochrone \citep{choi_mesa_2016}.  The statistical uncertainties are
propagated from the absolute magnitude (mostly originating from the
parallax uncertainty) and the color; the systematic uncertainties are
taken to be the difference between the PARSEC
\citep{bressan_parsec_2012} and MIST isochrones.  Reported
uncertainties are a quadrature sum of the statistical and systematic
components.  As a consistency check, we analyzed the aforementioned
Keck/HIRES spectrum from the night of 2021 March 26 using a
combination of \texttt{SpecMatch-Emp} for stellar properties, and
\texttt{SpecMatch-Synth} for $v\sin i$ \citep{yee_SM_2017}.  This
procedure yielded $T_{\rm eff}=5498\pm100\,{\rm K}$, $\log
g=4.6\pm0.1$, $[{\rm Fe/H}]=0.15\pm0.10$ from \texttt{SpecMatch-Emp},
and $v\sin i = 18.9\pm1.0$ from \texttt{SpecMatch-Synth}.  These
values are within the $1$-$\sigma$ uncertainties of our adopted values
from the isochrone interpolation.

\subsection{Kepler\,1627B}

We first noted the presence of a close neighbor in the \sn\ system on
2015 July 22 when we acquired adaptive optics imaging using the NIRC2
imager on Keck-II.  We used the narrow camera (FOV = 10.2\arcsec) to
obtain 8 images in the $K'$ filter ($\lambda = 2.12\,\mu$m) with a
total exposure time of 160\,s. We analyzed these data following
\citet{kraus_impact_2016}, which entailed using PSF-fitting to measure
the separation, position angle, and contrast of the candidate
companion.  The best-fitting empirical PSF template was identified
from among the near-contemporaneous observations of single stars in
the same filter.  The mean values inferred from the 8 images are
reported in Table~\ref{tab:starparams}.  To estimate the detection
limits, we analyzed the residuals after subtracting the empirical PSF
template. Within each residual image, the flux was measured through
40\,mas apertures centered on every pixel, and then the noise as a
function of radius was estimated from the RMS within concentric rings.
Finally, the detection limits were estimated from the strehl-weighted
sum of the detection significances in the image stack, and we adopted
the $6$-$\sigma$ threshold as the detection limit for ruling out
additional companions.

We also observed \sn\ on Gemini-North using the `Alopeke speckle
imager on 2021 June 24.  `Alopeke is a dual-channel speckle
interferometer that uses narrow-band filters centered at 0.83\,$\mu$m
and $0.56\,\mu$m.  We acquired three sets of $1000\times 60$$\,$msec
exposures during good seeing (0.45$''$), and used the autocorrelation
function of these images to reconstruct a single image and 5-$\sigma$
detection limits (see \citealt{howell_speckle_2011}).  This procedure
yielded a detection of the companion in the 0.83\,$\mu$m notch filter,
but not the $0.56\,\mu$m filter.  The measured projected separation
and magnitude difference are given in Table~\ref{tab:starparams}.

Figure~\ref{fig:kep1627binary} summarizes the results of the
high-resolution imaging.  The Gaia EDR3 parallax for the primary
implies a projected separation of $53 \pm 4$\,AU, assuming the
companion is bound.  Although the companion is unresolved in the Gaia
source catalog (there are no comoving, codistant candidate companions
brighter than $G < 20.5$ mag within $\rho < 120\arcsec$), its
existence was also suggested by the primary star's large \replaced{renormalized
unit weight error (RUWE),}{RUWE} relative to other members of the \cn\
\added{(RUWE$\approx$2.9; roughly the $98^{\rm th}$ percentile of the cluster's
distribution)}.  Based
on the apparent separation, the binary orbital period is of order
hundreds of years.  The large RUWE is therefore more likely to be
caused by a PSF-mismatch skewing the Gaia centroiding during
successive scans, rather than true astrometric motion.  Regardless,
given the low geometric probability that a companion imaged at $\rho
\approx 0\farcs16$ is a chance line-of-sight companion, we proceed
under the assumption that the companion is bound, and that
Kepler\,1627 is a binary.  Given the distance and age, the models of
\citet{baraffe_new_2015} imply a companion mass of \replaced{$M_{\rm B} \approx
0.33 M_{\odot}$}{$M_{\rm B} \approx 0.30\,M_{\odot}$} and companion
temperature of \replaced{$T_{\rm eff,B} \approx 3450$}{$T_{\rm eff,B}
\approx 3408$}\,K.  The corresponding spectral type is roughly \replaced{M2.5}{M3}V
\citep{pecaut_mamajek_2013}.  These models combined with the NIRC2
contrast limits imply physical limits on tertiary companions of
$M_{\rm ter} < 50 M_{\rm Jup}$ at $\rho = 50$\,AU, $M_{\rm ter} < 20
M_{\rm Jup}$ at $\rho = 100$\,AU, and $M_{\rm ter} < 10 M_{\rm Jup}$
at $\rho = 330$\,AU.

\section{The Planet}
\label{sec:planet}

\subsection{Kepler Light Curve}

\begin{figure*}[tp]
	\begin{center}
		\leavevmode
		\subfloat{
			\includegraphics[width=\textwidth]{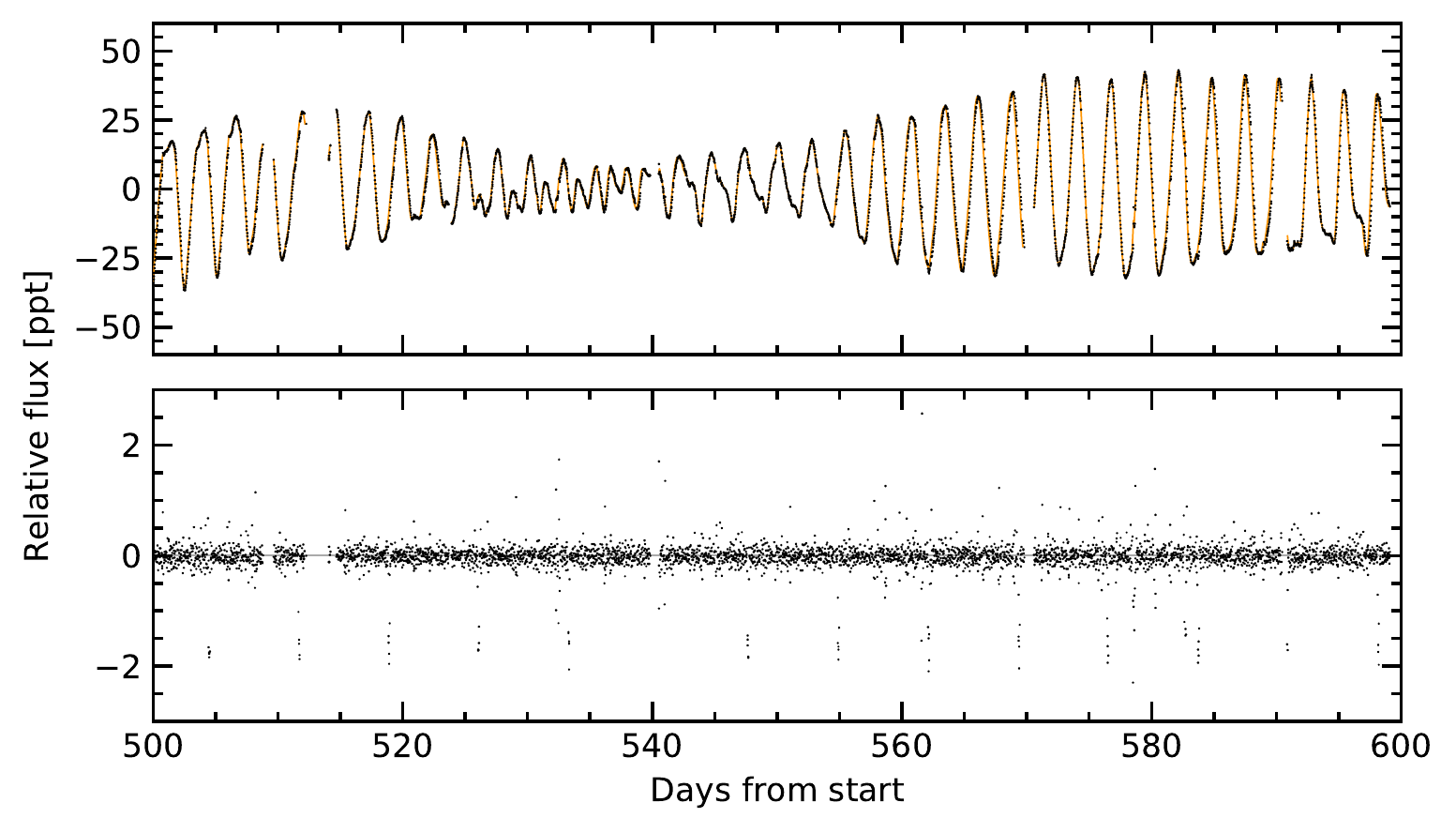}
		}

		\vspace{-0.2cm}	
		\subfloat{
			\includegraphics[width=0.5\textwidth]{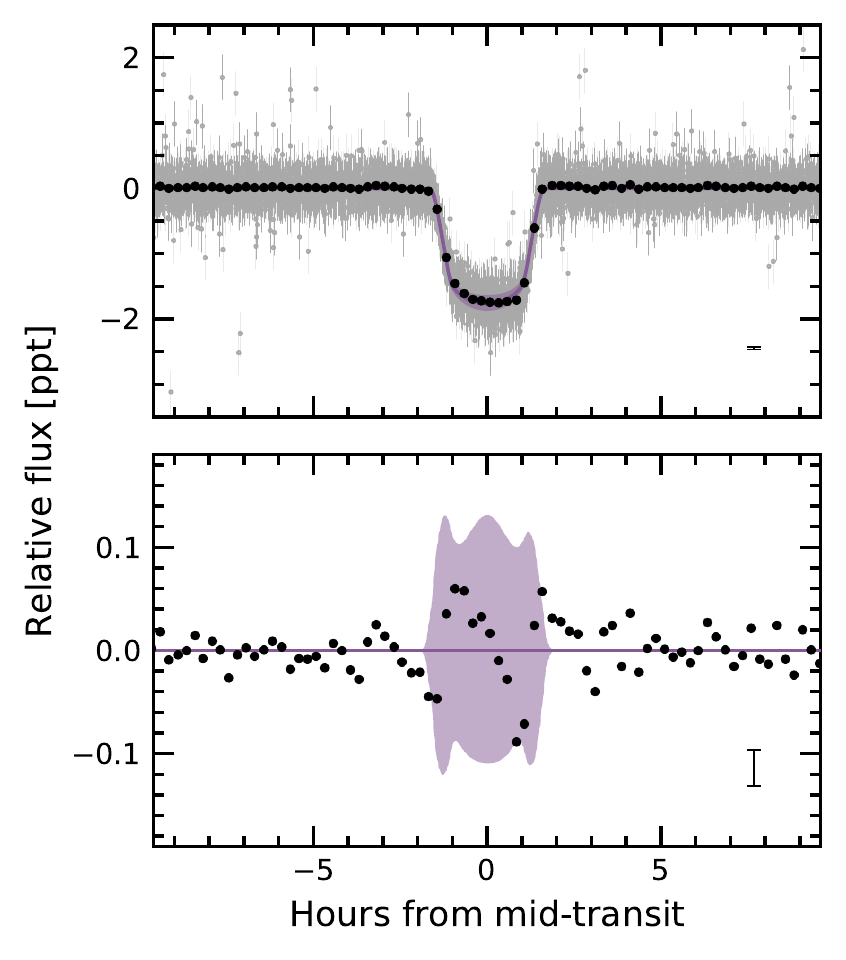}
			\includegraphics[width=0.5\textwidth]{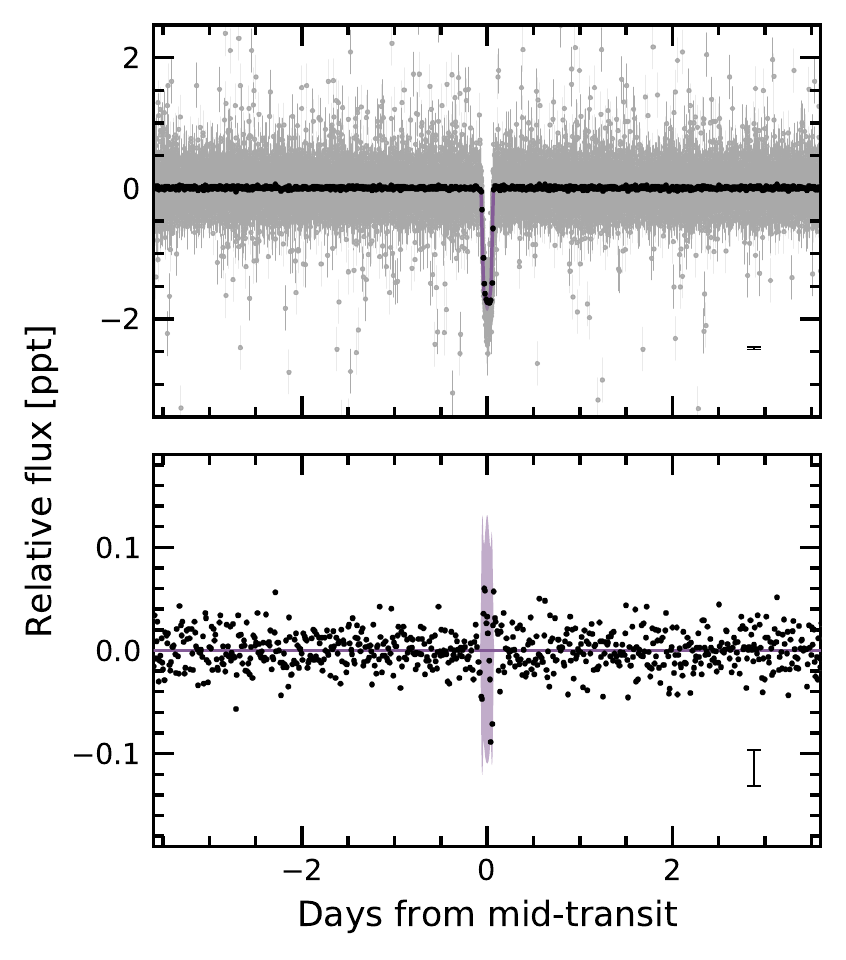}
		}
	\end{center}
	\vspace{-0.7cm}
  \caption{ {\bf The light curve of Kepler\,1627.}
    {\it Top}: 
    The Kepler data span 1{,}437 days (3.9 years), sampled at 30
    minute cadence;  a 100 day segment is shown.  The top panel shows
    the \texttt{PDCSAP} median-subtracted flux in units of
    parts-per-thousand ($\times 10^{-3}$).  The dominant signal is
    induced by starspots.  The stellar variability model (orange line)
    is subtracted below, revealing the transits of \pn.  The online
    Figure Set spans the entire 3.9 years of observations.
    {\it Bottom}:
    Phase-folded transit of Kepler 1627Ab with stellar variability
    removed.  Windows over 20 hours ({\it left}) and the entire orbit
    ({\it right}) are shown, and the residual after subtracting the
    transit is in the bottom-most row.  The 2-$\sigma$ model uncertainties
    and the best-fit model are the light purple band and the dark
    purple line.  Gray points are individual flux measurements; black
    points bin these to 15 minute intervals, and have a representative
    1-$\sigma$ error bar in the lower right of each panel.  The asymmetric
    residual during transit is larger than the out-of-transit scatter.
    \label{fig:lc}
  }
\end{figure*}

The Kepler space telescope observed \sn\ at a 30-minute cadence from
2009 May 2 until 2013 April 8.  Data gaps during quarters 4, 9, and 13
led to an average duty cycle over the 3.9~year interval of 67\%.  \sn\
was also observed at 1-minute cadence from 2012 Oct 5 until 2013 Jan
11.  The top panel of Figure~\ref{fig:lc} shows a portion of the
30-minute cadence \texttt{PDCSAP} light curve.  Nonastrophysical
variability has been removed using the methods discussed by
\citet{smith_kepler_PDC_2017}; the default optimal aperture was
assumed \citep{smith_finding_2016}.  Cadences with non-zero quality
flags (9\% of the data) have been omitted.  The resulting photometry
is dominated by a quasi-periodic starspot signal with a peak-to-peak
amplitude that varies between 2\% and 8\%.  \added{Given that the secondary
companion's brightness in the Kepler band is $\approx$1.5\% that of the primary,
source confusion for the rotation signal is not expected to be
an issue.}
Previous analyses have
identified and characterized the smaller transit signal
\citep{2012ApJS..199...24T,thompson_planetary_2018}, validated its
planetary nature \citep{morton_false_2016}, and even searched the
system for transit timing variations \citep{holczer_transit_2016}.
Nonetheless, since the cluster membership provides us with more
precise stellar parameters than those previously available, we opted
to reanalyze the light curve.

\subsubsection{Transit and Stellar Variability Model}
We fitted the Kepler long cadence time series with a model that
simultaneously included the planetary transit and the stellar
variability.  The stellar variability was modeled with the
\texttt{RotationTerm} Gaussian Process kernel in \texttt{exoplanet}
\citep{exoplanet:exoplanet}.  This kernel assumes that the variability
is generated by a mixture of two damped simple harmonic oscillators
with characteristic frequencies set by 1/$P_{\rm rot}$ and its first
harmonic.  We additionally included a jitter term to inflate the flux
uncertainties in a manner that accounted for otherwise unmodeled
excess white noise, and let the eccentricity float.  For the
limb-darkening, we assumed a quadratic law, and sampled using the
uninformative prior suggested by \citet{exoplanet:kipping13}.

Our model therefore included 10 free parameters for the transit ($\{P,
t_0, \delta, b, u_1 ,u_2 ,R_\star, \log g, e, \omega \}$), 2 free
parameters for the light curve normalization and a white noise jitter
($\{\langle f \rangle, \sigma_f \}$), and 5 hyperparameters for the GP
($\{\sigma_{\mathrm{rot}}, P_{\mathrm{rot}}, Q_0, \mathrm{d}Q, f \}$).
We also considered including an additive \texttt{SHOTerm} kernel to
account for stochastic noise, but found that this did not affect the
results, and so opted for the simpler GP kernel.  We fitted the models
using \texttt{PyMC3} \citep{salvatier_2016_PyMC3,exoplanet:theano},
and accounted for the finite integration time of each exposure in the
numerical integration when evaluating the model light curve
\citep[see][]{kipping_binning_2010}.  We assumed a Gaussian
likelihood, and after initializing each model with the parameters of
the maximum {\it a posteriori} model, we sampled using
\texttt{PyMC3}'s gradient-based No-U-Turn Sampler
\citep{hoffman_no-u-turn_2014} in the bases indicated in
Table~\ref{tab:posterior}.  We used $\hat{R}$ as our convergence
diagnostic \citep{gelman_inference_1992}.

Figure~\ref{fig:lc} shows the resulting best-fit model in orange (top)
and purple (bottom).  The model parameters and their uncertainties,
given in Table~\ref{tab:posterior}, are broadly consistent with a
\replaced{mini-Neptune }{Neptune-}sized planet ($3.82\pm0.16\,R_\oplus$) on a close-in
circular\footnote{ Our transit fitting yields $e<0.48$ at 2-$\sigma$;
the constraints on the eccentricity are not particularly strong.}
orbit around a G8V host star ($0.88 \pm 0.02 R_\odot$).  This best-fit
planet size is consistent with those previously reported by
\citet{morton_false_2016} and \citet{berger_2018_radii}, and
corrects for the small amount of flux dilution from Kepler\,1627B.

\subsubsection{Transit Asymmetry}
\label{subsec:asymmetry}
The transit fit however is not perfect: the lower panels of
Figure~\ref{fig:lc} show an asymmetric residual in the data relative
to the model: the measured flux is high during the first half of
transit, and low in the second half.  The semi-amplitude of this
deviation is $\approx50\,{\rm ppm}$, which represents a $\approx 3\%$
distortion of the transit depth ($\delta=1759\pm62\,{\rm ppm}$).  Note that
although this asymmetry is within the 2-$\sigma$ model uncertainties, the
model has a jitter term that grows to account for excess
white noise in the flux.  The significance of the asymmetry is
therefore best assessed in comparison against the intrinsic
out-of-transit scatter in the data ($\approx16\,{\rm ppm}$), not the model uncertainties.  The
lower right panel of Figure~\ref{fig:lc} demonstrates that the scatter during
transit is higher than during all other phases of the planet's orbit.

To determine whether the asymmetry could be a systematic caused by our
stellar variability model, we explored an alternative approach in
which we isolated each transit window, locally fitted out
polynomial trends, and then binned all the observed transits; the
asymmetry was still present at a comparable amplitude.
Appendix~\ref{app:asymmetry} describes a more detailed analysis, which finds
that the asymmetry also seems to be robust to different methods of
data binning in time and by local light curve slope.  Possible
astrophysical explanations are discussed in Section~\ref{sec:conc}.

\subsubsection{Transit Timing and the Local Slope}
\label{subsec:ttvslope}

\begin{figure}[tp]
	\begin{center}
		\leavevmode
		\includegraphics[width=0.47\textwidth]{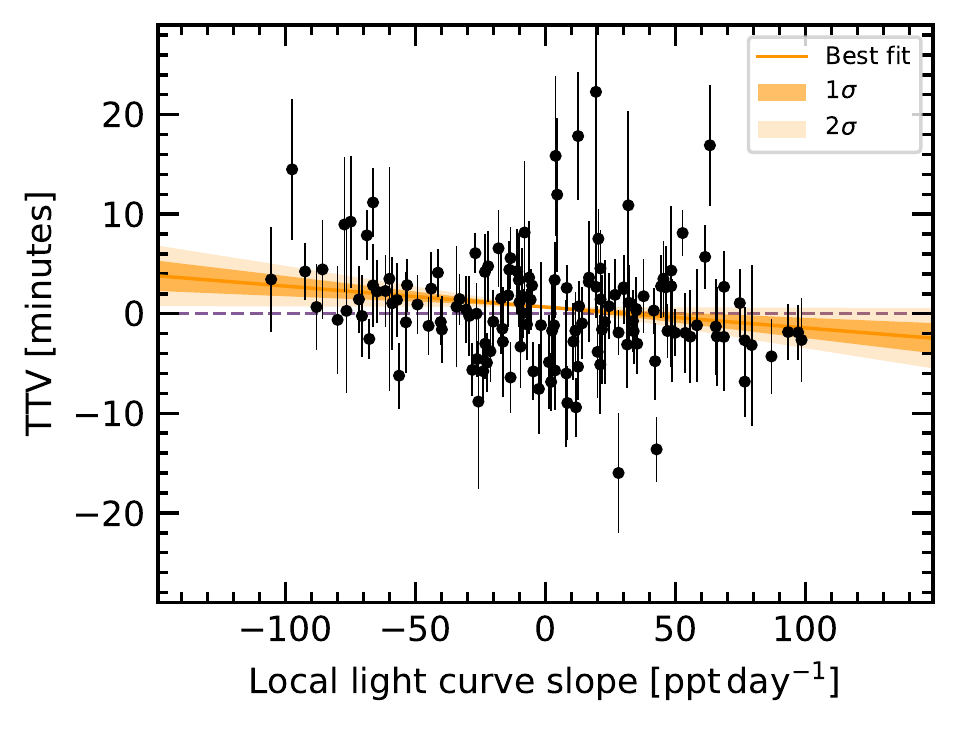}
	\end{center}
	\vspace{-0.6cm}
	\caption{
    {\bf \replaced{Possible}{Weak} evidence for a prograde orbit of Kepler 1627 Ab.} The time of
    each Kepler transit was measured, along with the local slope of
    the light curve.  The two quantities \replaced{are weakly}{might
    be} anti-correlated
    ($\approx$2-$\sigma$), which
    \replaced{might}{could} be caused by starspot crossings during the first
    (second) half of transit inducing a positive (negative) TTV,
    provided that the orbit is prograde \citep{mazeh_time_2015}.  The
    units along the abscissa can be understood by considering that the
    stellar flux changes by $\sim$60\,ppt per half rotation period
    ($\sim$1.3\,days).
		\label{fig:corr}
	}
\end{figure}

The previous analysis by \citet{holczer_transit_2016} did not
find any significant long-term transit timing or duration variations
(TTVs or TDVs) for \sn.  Quantitatively, the mean and standard deviation of the
TTVs and TDVs they measured were $-1.1\pm13.8\,{\rm
min}$ and $-3.3\pm22.1\,{\rm min}$.  In an
earlier analysis however, \citet{holczer_time_2015} studied correlations
between TTVs and local light curve slopes, and for \sn\
found a weak correlation of $-29\pm13\,{\rm min}\,{\rm day}^{-1}$
between the two quantities.  Given the possible connection between such
correlations and the unresolved starspot crossings that we expect to
be present in the \sn\ light curve \citep{mazeh_time_2015}, we opted to re-examine the
individual transit times.

We therefore isolated each of the 144 observed transits to within
$\pm4.5$\,hr of each transit, and fitted each window with both {\it
i)} a local \added{second or fourth-order }polynomial baseline plus the transit, and {\it ii)} a local linear
trend\added{ plus the transit}.  \deleted{We considered the results both for a second and fourth-order
time-dependence in the local baseline.  }We let the mid-time of each
transit float, and then calculated the residual between the measured
mid-time and that of a periodic orbit.  This residual, the transit
timing variation, is plotted in Figure~\ref{fig:corr} against
the local linear slope \replaced{for}{assuming} the fourth-order polynomial baseline.  The
slope of $-21 \pm 10\,{\rm min\,day}^{-1}$ is similar to that
found by \citet{holczer_time_2015}. \added{The
best-fit line yields $\chi^2 = 306.1$, with $n=140$ data points.  An alternative
model of just a flat line yields $\chi^2=315.6$.  The difference in the
Bayesian information criterion between the two models is ${\rm
BIC}_{\rm flat} - {\rm BIC}_{\rm line} = 4.5$,
which corresponds to a Bayes factor of $\approx$9.4. According to the
usual \citet{kass_bayes_1995} criteria, this is ``positive'' evidence
for the model with a finite slope.  We view it as suggestive
at best, particularly given the poor reduced $\chi^2$.}

\replaced{One}{A separate} concern we had in this analysis was whether our transit fitting
procedure might induce spurious correlations between the slope and
transit time.  In particular, assuming the second-order polynomial
baseline yielded a larger anti-correlation between the TTVs and local
slopes, of $-79 \pm 14\,{\rm min\,day}^{-1}$.  We therefore
performed an injection-recovery procedure in which we injected
transits at different phases in the \sn\ light curve and repeated the
TTV analysis.  This was done at $\approx$50 phases, each separated by
$0.02\,P_{\rm orb}$ while omitting the phases in transit.  For
the second-order polynomial baseline, this procedure yielded
a similar anti-correlation in the injected transits as that present in the real
transit;
assuming this baseline would therefore bias the
result.  However for the fourth-order baseline, the correlation
present in the data was stronger than in all but one of the injected
transits.
Possible interpretations are discussed below. Given the
\added{lack of }statistical significance, this analysis should be interpreted as
suggestive at best.

\subsection{Planet Confirmation}

\begin{figure}[tp]
	\begin{center}
		\leavevmode
		\includegraphics[width=0.47\textwidth]{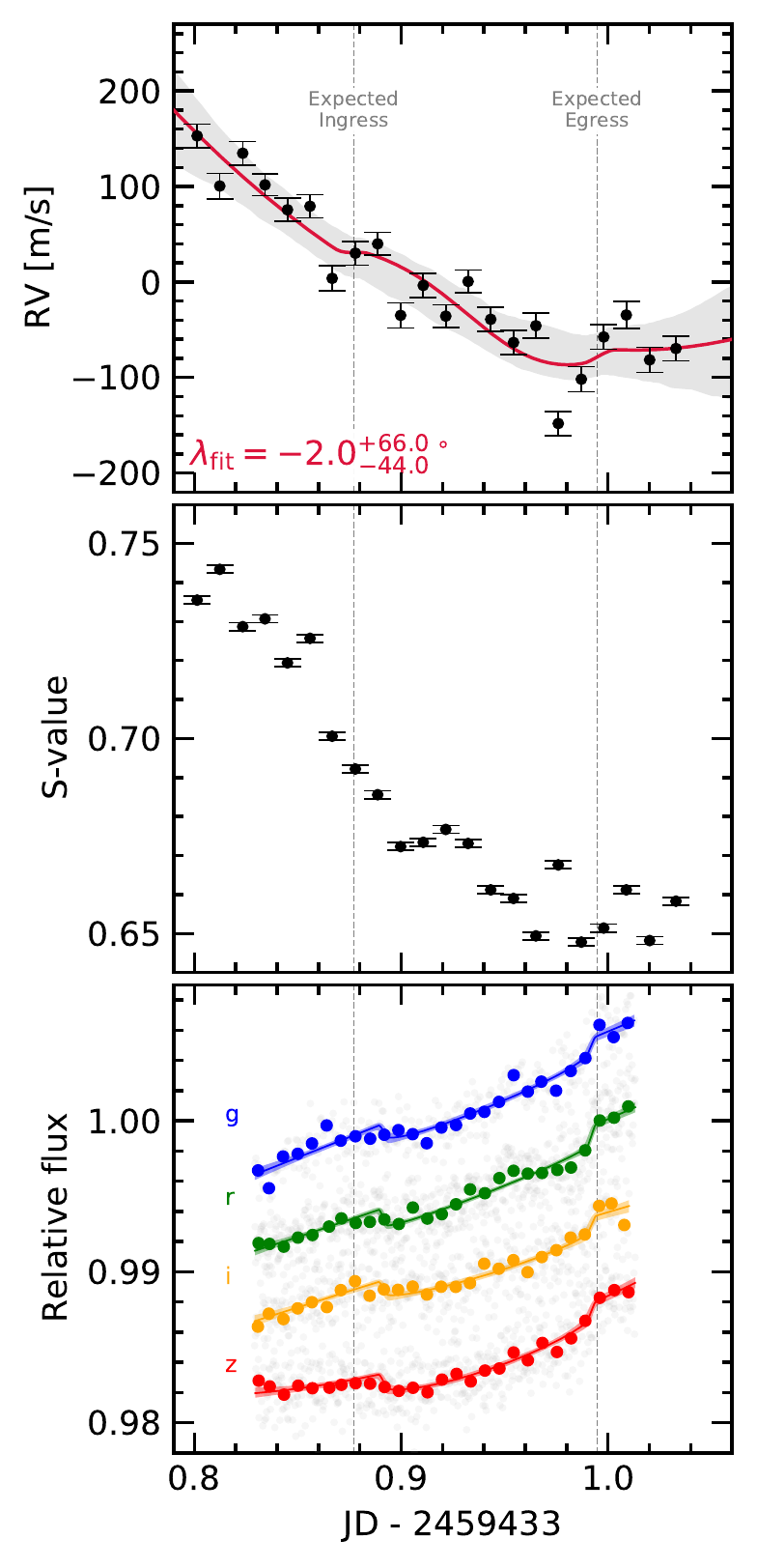}
	\end{center}
	\vspace{-0.7cm}
	\caption{
		{\bf Keck/HIRES radial velocities and MuSCAT3 photometry from the transit of 2021 Aug 7.}
		{\it Top:} 
		The radial velocity jitter across the 15~minute exposures
		($\sigma_{\rm RV}\approx30$\ms) prevented us from detecting the RM
		effect;
		a model including the RM anomaly and a quadratic trend in time to
		fit the spot-induced $\approx$\,$250$\ms\ trend is shown (see
		Appendix~\ref{app:groundobs}).  Shaded bands show 2-$\sigma$ model
		uncertainties.  
		{\it Middle:} The RV variations are strongly correlated with varying
		emission in the Ca H and K lines.
		{\it Bottom}: The photometric transit depths are consistent across
		the {\it griz} bandpasses. The photometry is binned at 10~minute
		intervals.
		\label{fig:ground}
	}
\end{figure}

If the \pn\ transit signal is created by a genuine planet,  then to
our knowledge it would be the youngest planet yet found by the
\replaced{main}{prime} Kepler mission.\footnote{The re-purposed K2 mission however has found
two younger systems containing five planets: K2-33b ($9\pm1\,{\rm
Myr}$; \citealt{Mann_K2_33b_2016,David_et_al_2017}) and V1298 Tau
($23\pm4\,{\rm Myr}$; \citealt{david_four_2019}).}   Could the transit
be produced by anything other than a planet orbiting this near-solar
analog?  \citet{morton_false_2016} validated the planet based on the
transit shape, arguing that the most probable false positive scenario
was that of a background eclipsing binary, which had a model-dependent
probability of $\approx$$10^{-5}$.  However, this calculation was
performed without knowledge of the low-mass stellar companion ($M_{\rm
B}\approx 0.30\,M_\odot$).  Validated planets have also previously
been refuted \citep[{\it e.g.},][]{shporer_three_2017}.  We therefore
reassessed false positive scenarios in some detail. 


As an initial plausibility check, Kepler\,1627B contributes 1\% to 2\%
of the total flux observed in the Kepler aperture.  For the sake of
argument, assume the former value.  The observed transit has a depth
of $\approx$0.18\%.  An $\approx$18\% deep eclipse of Kepler\,1627B would
therefore be needed to produce a signal with the appropriate depth.
The shape of the transit signal however requires the impact parameter
to be below 0.77 (2-$\sigma$); the tertiary transiting the secondary
would therefore need to be non-grazing with $R_3/R_2 \approx 0.4$.
This yields a contradiction:  this scenario requires an ingress
and egress phase that each span $\approx$40\% of the transit duration
($\approx 68\,{\rm min}$).  The actual measured ingress and egress
duration is $\approx$$17\,{\rm min}$, a factor of four times too
short.  The combination of Kepler 1627B's brightness, the transit
depth, and the ingress duration therefore disfavor the scenario that
Kepler 1627B might host the transit signal.  

Beyond this simple test, a line of evidence that
effectively confirms the planetary interpretation is that the stellar density
implied by the transit duration and orbital period is inconsistent
with an eclipsing body around the M-dwarf companion.  We find
$\rho_\star = 2.00 \pm 0.24\,{\rm g\,cm}^{-3}$, while the
theoretically expected density for Kepler\,1627B \added{given its
nominal age of 38\,Myr and mass of 0.30\,$M_\odot$} is $\approx$$4.6\,{\rm
g\,cm}^{-3}$ \citep{baraffe_new_2015}.
The transit duration is therefore too long to be explained by a star
eclipsing the M dwarf secondary at $10$-$\sigma$.\added{ Adversarially assuming a younger age (32\,Myr) and a larger companion
mass (0.35\,$M_\odot$), the companion's density could be as low as $\approx 3.5\,{\rm
g\,cm}^{-3}$.  This is still in tension with the fitted stellar density.} While the planet
might hypothetically still orbit a hidden close and bright companion,
this possibility is implausible given {\it i)} the lack of secondary
lines in the HIRES spectra, {\it ii)} the lack of secondary
rotation signals in the Kepler photometry, and {\it iii)} the proximity
of \sn\ to the \cn\ locus on the Gaia CAMD (Figure~\ref{fig:age}).


The correlation noted in Section~\ref{subsec:ttvslope} between the
TTVs and the local light curve slope might be an additional line of evidence
in support of the planetary interpretation.  Unless it is a
statistical fluke (a $\approx$5\% possibility), then the most likely
cause of the correlation is unresolved starspot
crossings \citep{mazeh_time_2015}.  These would only be possible if the
planet transits the primary star, which excludes a background
eclipsing binary scenario.  The correlation would also suggest that the planet's
orbit is prograde.  The latter point assumes that the dominant
photometric variability is induced by dark spots, and not bright
faculae.  Given the observed transition of Sun-like stellar
variability from spot to faculae-dominated regimes between young and
old ages, we expect this latter assumption to be reasonably secure
\citep{shapiro_are_2016,montet_long-term_2017,reinhold_stellar_2020}.

A third supporting line of evidence for the planetary interpretation
also exists.  We observed a transit of \pn\ on the
night of 2021 Aug 7 \replaced{simultaneously with Keck/HIRES and
MuSCAT3}{spectroscopically with HIRES at the Keck-I telescope and photometrically in
$griz$ bands with MuSCAT3 at Haleakal\=a Observatory}.  \added{
Details of the observation sequence are discussed in
Appendix~\ref{app:groundobs}; Figure~\ref{fig:ground} shows the
results.}\deleted{.  We
scheduled the observations using the ephemeris of
\citet{holczer_transit_2016}.}  Although we did not detect the
Rossiter-McLaughlin (RM) anomaly, the multi-band MuSCAT3 light curves
show that the transit is achromatic\deleted{
  (Figure~\ref{fig:ground})}.
Quantitatively, when we fitted the MuSCAT3
photometry with a model that lets the transit depths vary across each
bandpass, we found {\it griz} depths consistent with the Kepler depth
at 0.6, 0.3, 0.3, and 1.1-$\sigma$ respectively.  \added{The
achromatic transits strongly favor Kepler\ 1627A as the transit host,
since Kepler\ 1627B is a much redder star.}  Conditioned on the
ephemeris and transit depth from the Kepler data, the MuSCAT3 observations also
suggested a transit duration $17.3\pm4.3\ {\rm min}$ shorter than the
Kepler transits.
However, given both the lack of TDVs in the Kepler data
and the relatively low signal-to-noise of the MuSCAT3 transit,
further photometric follow-up would be necessary to determine
whether the transit duration is \replaced{indeed}{actually} changing.  

For our RM analysis,
the details are discussed in Appendix~\ref{app:groundobs}.  While the
velocities are marginally more consistent with a prograde or polar
orbit than a retrograde orbit, the spot-corrected exposure-to-exposure
scatter ($\sigma_{\rm RV}\approx 30$\ms) is comparable to the expected
RM anomaly assuming an aligned orbit ($\Delta v_{\rm RM}\approx
20\,{\rm m\,s}^{-1}$).  We are therefore not in a position to claim a
spectroscopic detection of the RM effect, nor to quantify the stellar
obliquity.

\section{Discussion \& Conclusions}
\label{sec:conc}

\begin{figure*}[!t]
	\begin{center}
		\leavevmode
		\includegraphics[width=0.9\textwidth]{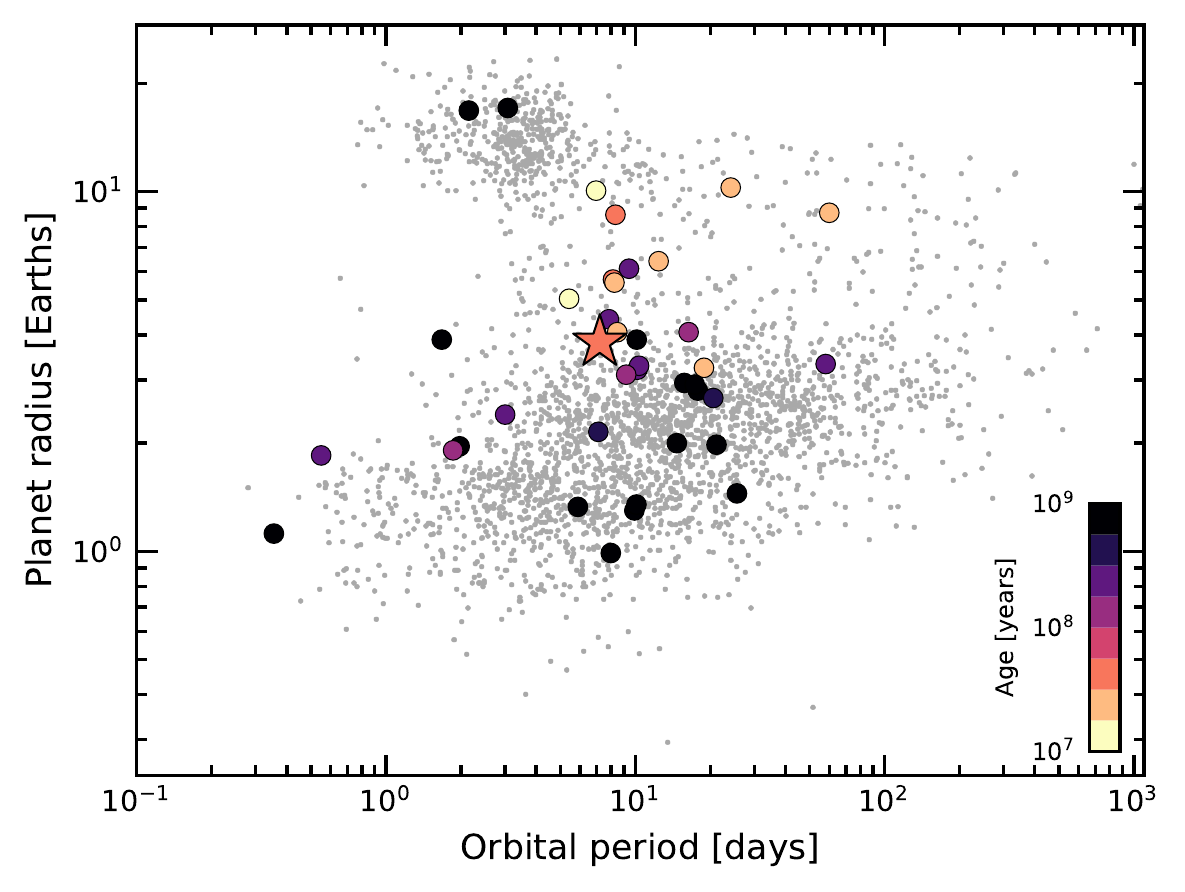}
	\end{center}
	\vspace{-0.7cm}
	\caption{
		{\bf Radii, orbital periods, and ages of transiting exoplanets}.
		Planets younger than a gigayear with ${\rm \tau}/\sigma_{\tau} >
		3$ are emphasized, where $\tau$ is the age and $\sigma_{\tau}$ is
		its uncertainty.  \pn\ is shown with a star.  The large sizes of
		the youngest transiting planets could be explained by their
		primordial atmospheres not yet having evaporated; direct
		measurements of the atmospheric outflows or planetary masses would
		help to confirm this expectation.  Selection effects may also be
		important.  Parameters are from the NASA Exoplanet Archive (2021
		Sept 15).
		\label{fig:rp_period_age}
	}
\end{figure*}

Kepler\,1627Ab provides a new extremum in the ages of the Kepler
planets, and opens multiple avenues for further study.  Observations
of spectroscopic transits at greater
\replaced{precision}{signal-to-noise} should yield a
measurement of the stellar obliquity, which would confirm or refute
the prograde orbital geometry suggested by the TTV-local slope
correlation.  Separately, transit spectroscopy aimed at detecting
atmospheric outflows could yield insight into the evolutionary state
of the atmosphere \citep[{\it
e.g.},][]{ehrenreich_giant_2015,spake_helium_2018,vissapragada_2020}.
Observations that quantify the amount of high-energy
irradiation incident on the planet would complement these efforts, by
helping to clarify the expected outflow rate \citep[{\it
e.g.},][]{poppenhaeger_2021}.  Finally, a challenging but informative
quantity to measure would be the planet's mass.  Measured at
sufficient precision,
for instance through a multi-wavelength radial velocity campaign,
the combination of the size, mass, and age 
would yield constraints on both the planet's composition and its
initial entropy \citep{owen_constraining_2020}.

More immediately, the Kepler data may yet contain additional
information.  For instance, one possible explanation for the transit
asymmetry shown in Figure~\ref{fig:lc} is that of a dusty asymmetric
outflow.  
Dusty outflows are theoretically
expected for young mini-Neptunes, and the amplitude of the observed
asymmetry is consistent with predictions \citep{wang_dai_2019}.
A second possibility is that the planetary orbit is slightly
misaligned from the stellar spin axis, and tends to transit starspot
groups at favored stellar latitudes.  
This geometry would be necessary in order to
explain how the starspot crossings could add up coherently\added{,
given that the planetary orbital period (7.203\,days) and the stellar
rotation period (2.642\,days) are not a rational combination}.
Other possibilities including
gravity darkening or TTVs causing the asymmetry are disfavored (see
Appendix~\ref{app:asymmetry}).  

Beyond the asymmetric transits, Appendix~\ref{app:flare}
highlights an additional abnormality in the short-cadence Kepler data, in the
arrival time distribution of stellar flares.  We encourage its
exploration by investigators more versed in the topic than ourselves.

In the context of the transiting planet population, \pn\ is among the
youngest known (Figure~\ref{fig:rp_period_age}).  Comparable systems
with precise ages include K2-33
\citep{Mann_K2_33b_2016,David_et_al_2017}, DS~Tuc
\citep{benatti_possibly_2019,newton_tess_2019}, HIP~67522
\citep{rizzuto_tess_2020}, TOI~837 \citep{bouma_cluster_2020}, the
two-planet AU~Mic \citep{plavchan_planet_2020,martioli_aumicbc_2021}
and the four-planet V1298~Tau \citep{david_four_2019}.  \pn\ is one of
the smaller planets in this sample ($3.82\pm0.16\,R_\oplus$), which
could be linked to the selection effects imposed by spot-induced
photometric variability at very young ages \citep[{\it
e.g.},][]{zhou_2021_tois}. 

 \replaced{However,}{It seems though that} smaller planets could have been
detected\added{ in the \sn\ system}: \added{based on the per-target detection contours, }the Kepler pipeline's median completeness extended to
$1.6\,R_\oplus$ at 10~day orbital periods, and $3.3\,R_\oplus$ at
100~days
\citep{2021ascl.soft07027B}.
\added{These limits account for the spot-induced variability in the
system through
a correction based on the Combined Differential Photometric Precision
in the light curve over the relevant transit timescales
\citep{2017ksci.rept...19B}.} The large size of \pn\ relative
to most Kepler mini-Neptunes might therefore support a picture
in which the typical $5\,M_\oplus$\deleted{ to 10$\,M_\oplus$} mini-Neptune
\citep{wu_mass_2019} loses a significant fraction of its primordial
atmosphere over its first gigayear
\citep{Owen_Wu_2013,ginzburg_corepowered_2018}.  
\added{
It could also be consistent with a scenario in which an earlier ``boil-off'' of
the planet's atmosphere during disk dispersal decreases the entropy of
the planetary interior, leading to a rather long $\sim$10$^8$ year Kelvin-Helmholtz
contraction timescale
\citep{owen_constraining_2020}.
Confirming either of these scenarios would require a measurement
of the planetary mass; otherwise, alternative explanations for its large size
could include that it is just abnormally massive, or that it has an abnormally
large envelope to core mass ratio.}


Ultimately, the main advance of this work is a precise measurement of
the age of \pn.   This measurement was enabled by identifying the
connection of the star to the \cn\ using Gaia kinematics, and by then 
using the Gaia color-absolute magnitude diagram and TESS stellar
rotation periods to verify the cluster's existence.
Table~\ref{tab:v06} enables similar cross-matches for both known and
forthcoming exoplanet systems \citep[{\it
e.g.},][]{guerrero_tess_2021}. Confirming these candidate associations
using independent age indicators is essential because their false
positive rates are not known.  A related path is to identify new
kinematic associations around known exoplanet host stars using
positions and tangential velocities from Gaia, and to verify
these associations with stellar rotation periods and spectroscopy
\citep[{\it e.g.},][]{tofflemire_tess_2021}.  Each approach seems
likely to expand the census of planets with precisely measured ages
over the coming years, which will help in deciphering the early stages
of exoplanet evolution.


\acknowledgements
\raggedbottom

The authors are grateful to J{.}~Winn, J{.}~Spake, A{.}~Howard, and
T{.}~David for illuminating discussions and suggestions, and to
R{.}~Kerr for providing us with the \citet{Kerr2021} membership list
prior to its publication.  The authors are also grateful to
K{.}~Collins for helping resolve the scheduling conflict that would
have otherwise prevented the MuSCAT3 observations.
L.G.B{.} acknowledges support from a Charlotte Elizabeth Procter
Fellowship from Princeton University, as well as from the TESS GI
Program (NASA grants 80NSSC19K0386 and 80NSSC19K1728) and the
Heising-Simons Foundation (51 Pegasi~b Fellowship).
%
%
Keck/NIRC2 imaging was acquired by program 2015A/N301N2L
(PI: A.~Kraus). 
In addition, this paper is based in part on observations made with the
MuSCAT3 instrument, developed by the Astrobiology Center and under
financial support by JSPS KAKENHI (JP18H05439) and JST PRESTO
(JPMJPR1775), at Faulkes Telescope North on Maui, HI, operated by the
Las Cumbres Observatory.
This work is partly supported by JSPS KAKENHI Grant Numbers 22000005,
JP15H02063, JP17H04574, JP18H05439, JP18H05442, JST PRESTO Grant
Number JPMJPR1775, the Astrobiology Center of National Institutes of
Natural Sciences (NINS) (Grant Number AB031010).
%
%
This paper also includes data collected by the TESS mission, which are
publicly available from the Mikulski Archive for Space Telescopes
(MAST).
Funding for the TESS mission is provided by NASA's Science Mission
directorate.
We thank the TESS Architects (G.~Ricker, R.~Vanderspek, D.~Latham,
S.~Seager, J.~Jenkins) and the many TESS team members for their
efforts to make the mission a continued success.
%
%
%
%
Finally, we also thank the Keck
Observatory staff for their support of HIRES and remote observing.  We
recognize the importance that the summit of Maunakea has within the
indigenous Hawaiian community, and are deeply grateful to have the
opportunity to conduct observations from this mountain.

\software{
  \texttt{altaipony} \citep{ilin_flares_2021},
  \texttt{astrobase} \citep{bhatti_astrobase_2018},
  \texttt{astropy} \citep{astropy_2018},
  \texttt{astroquery} \citep{astroquery_2018},
  \texttt{corner} \citep{corner_2016},
  \texttt{exoplanet} \citep{exoplanet:exoplanet}, and its
  dependencies \citep{exoplanet:agol20, exoplanet:kipping13, exoplanet:luger18,
   	exoplanet:theano},
  \texttt{PyMC3} \citep{salvatier_2016_PyMC3},
  \texttt{scipy} \citep{jones_scipy_2001},
  \texttt{TESS-point}  \citep{burke_2020},
  \texttt{wotan} \citep{hippke_wotan_2019}.
}
\ 

\facilities{
 	{\it Astrometry}:
 	Gaia \citep{gaia_collaboration_gaia_2018,gaia_collaboration_2021_edr3}.
 	{\it Imaging}:
    Second Generation Digitized Sky Survey. 
 	Keck:II~(NIRC2; \url{www2.keck.hawaii.edu/inst/nirc2}).
 	Gemini:North~(`Alopeke; \citealt{scott_nessi_2018,scott_twin_2021}.
 	{\it Spectroscopy}:
 	Keck:I~(HIRES; \citealt{vogt_hires_1994}).
 	{\it Photometry}:
	  Kepler \citep{borucki_kepler_2010},
    MuSCAT3 \citep{Narita_2020},
 	  TESS \citep{ricker_transiting_2015}.
}

\input{starparams.tex}
\input{model_posterior_table.tex}
\input{t1_appendix.tex}
\input{t2_appendix.tex}

\clearpage
\bibliographystyle{yahapj}                            
\bibliography{bibliography} 

\appendix
\section{Young, Age-Dated, and Age-Dateable Star Compilation}
\label{app:targetlist}

The \texttt{v0.6} CDIPS target catalog (Table~\ref{tab:v06}) includes
stars that are young, age-dated, and age-dateable.  By
``age-dateable'', we mean that the stellar age should be measurable at
greater precision than that of a typical FGK field star, through
either isochronal, gyrochronal, or spectroscopic techniques.  As in
\citet{bouma_cdipsI_2019}, we collected stars that met these criteria
from across the literature.  Table~\ref{tab:metadata} gives a list of
the studies included, and brief summary statistics.  The age
measurement methodologies adopted by each study differ: in many,
spatial and kinematic clustering has been performed on the Gaia data,
and ensemble isochrone fitting of the resulting clusters has been
performed (typically focusing on the turn-off).  In other studies
however, the claim of youth is based on the location of a single star
in the color-absolute magnitude diagram, or on spectroscopic
information.

One major change in Table~\ref{tab:v06} relative to the earlier
iteration from \citet{bouma_cdipsI_2019} is that the extent of
Gaia-based analyses has now matured to the point that we can neglect
pre-Gaia cluster memberships, except for a few cases with
spectroscopically confirmed samples of age-dated stars.  The
membership lists for instance of \citet{Kharchenko_et_al_2013} and
\citet{dias_proper_2014} (MWSC and DAML) are no longer required.  This
is helpful for various post-processing projects,  since the field star
contamination rates were typically much higher in these catalogs than
in the newer Gaia-based catalogs.

The most crucial parameters of a given star for our purposes are the
Gaia DR2 $\texttt{source\_id}$, the cluster or group name
($\texttt{cluster}$), and the $\texttt{age}$.  Given the hierarchical
nature of many stellar associations, we do not attempt to resolve the
cluster names to a single unique string.  The Orion complex for
instance can be divided into almost one hundred kinematic subgroups
\citep{kounkel_apogee2_2018}.  \added{Based on Figure~\ref{fig:XYZvtang}, the
\cn\ may also be part of a similar hierarchical association.} Similar complexity applies to the
problem of determining homogeneous ages, which we do not attempt to
resolve.  Instead, we simply merged the cluster names and ages
reported by various authors into a comma-separated string.

This means that the \texttt{age} column can be null, for cases in
which the original authors did not report an age, or for which a
reference literature age was not readily available.  Nonetheless,
since we do prefer stars with known ages, we made a few additional
efforts to populate this column.  When available, the age provenance
is from the original analysis of the cluster.  In a few cases however
we adopted other ages when string-based cross-matching on the cluster
name was straightforward.  In particular, we used the ages determined
by \citet{CantatGaudin2020b} to assign ages to the clusters from
\citet{GaiaCollaboration2018}, \citet{CantatGaudin2018a},
\citet{CastroGinard2020}, and \citet{CantatGaudin2020a}.

The catalogs we included for which ages were not immediately available
were those of \citet{CottenSong2016}, \citet{Oh2017},
\citet{Zari2018}, \citet{Gagne2018b}, \citet{Gagne2018a},
\citet{Gagne2018c}, and \citet{Ujjwal2020}.  While in principle the
moving group members discussed by
\citet{Gagne2018b,Gagne2018a,Gagne2018c} and \citet{Ujjwal2020} have
easily associated ages, our SIMBAD cross-match did not retain the
moving group identifiers given by those studies, which should
therefore be recovered using tools such as BANYAN
$\Sigma$ \citep{Gagne2018a}.
We also included the SIMBAD object identifiers $\texttt{TT*}$,
$\texttt{Y*O}, $\texttt{Y*?}, $\texttt{TT?}$, and $\texttt{pMS*}$.
Finally, we  included every star in the NASA Exoplanet Archive
planetary system ($\texttt{ps}$) table that had a Gaia identifier
available \citep{NASAExoArchive_ps_20210506}.  If the age had finite
uncertainties, we also included it, since stellar ages determined
through the combination of isochrone-fitting and transit-derived
stellar densities typically have higher precision than from isochrones
alone.

For any of the catalogs for which Gaia DR2 identifiers were not
available, we either followed the spatial (plus proper-motion)
cross-matching procedures described in \citet{bouma_cdipsI_2019}, or
else we pulled the Gaia DR2 source identifiers associated with the
catalog from SIMBAD.  We consequently opted to drop the
$\texttt{ext\_catalog\_name}$ and $\texttt{dist}$ columns maintained
in \citet{bouma_cdipsI_2019}, as these were only populated for a small
number of stars.  The technical manipulations for the merging,
cleaning, and joining were performed using $\texttt{pandas}$
\citep{mckinney-proc-scipy-2010}.  The eventual cross-match (using the
Gaia DR2 $\texttt{source\_id}$) against the Gaia DR2 archive was
performed asychronously on the Gaia archive
website.

\section{The Transit Asymmetry}
\label{app:asymmetry}

\begin{figure}[t]
	\begin{center}
		\leavevmode
		\includegraphics[width=0.64\textwidth]{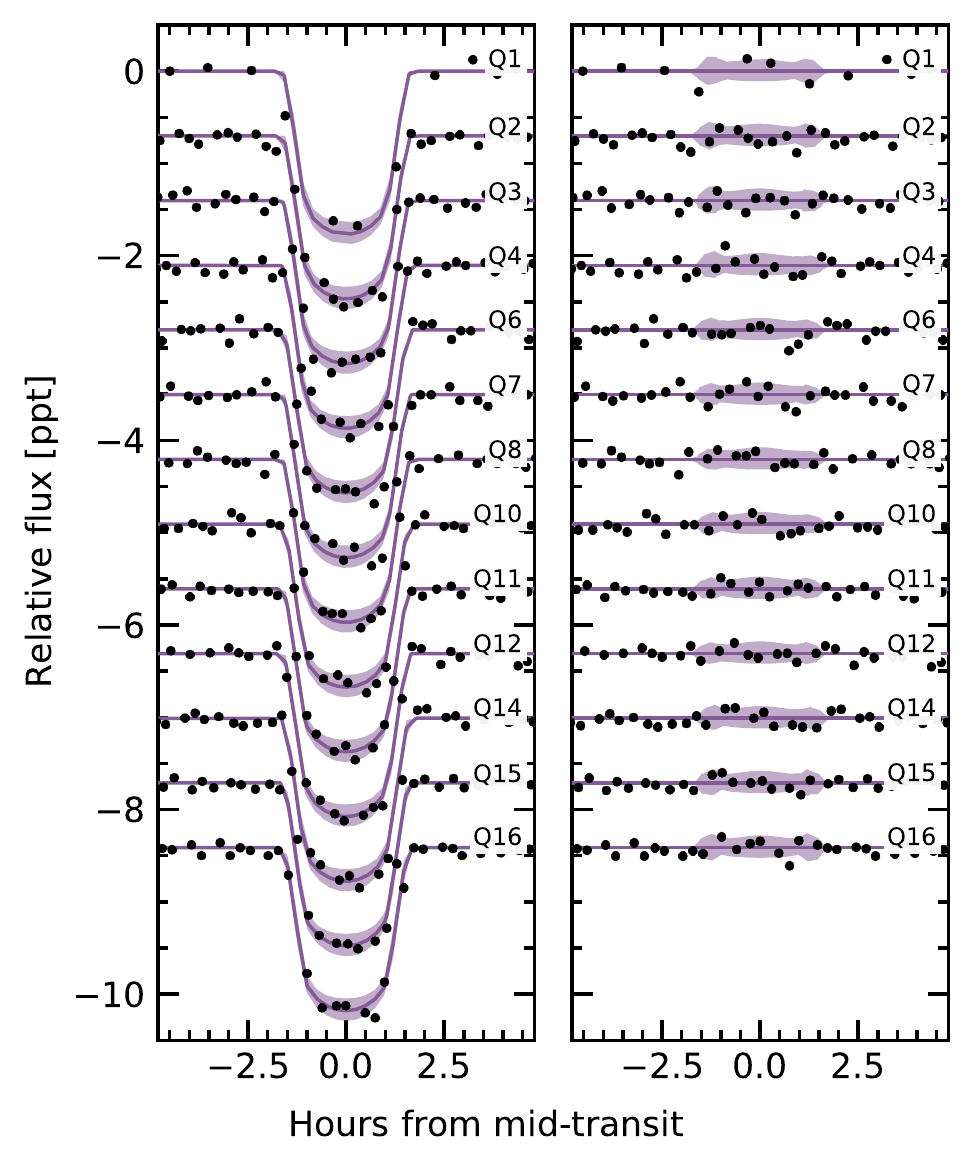}
	\end{center}
	\vspace{-0.7cm}
	\caption{
		{\bf Transit model residuals through time (binned by Kepler quarter)}.  
    {\it Left:}
    Phase-folded transit of Kepler 1627b, with stellar variability
    removed.  Black points are binned to 20 minute intervals.  The
    2-$\sigma$ model uncertainties and the maximum {\it a posteriori}
    model are shown as the faint purple band, and the dark purple
    line.
    {\it Right:}
    As on the left, with the transit removed.  
		\label{fig:phasequarter}
	}
\end{figure}

\begin{figure}[t]
	\begin{center}
		\leavevmode
		\includegraphics[width=0.64\textwidth]{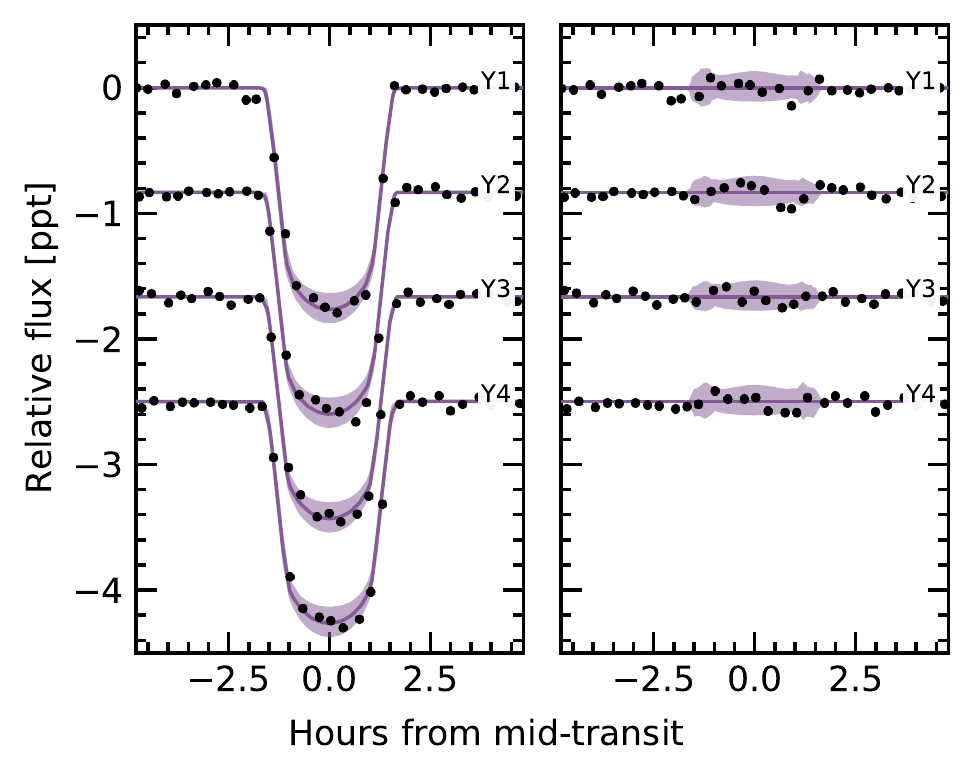}
	\end{center}
	\vspace{-0.7cm}
	\caption{
    {\bf Transit model residuals through time (binned by year of observation)}.  
    {\it Left:}
    Phase-folded transit of Kepler 1627b, with stellar variability
    removed.  Points and models are as in
    Figure~\ref{fig:phasequarter}.
    {\it Right:}
    As on the left, with the transit removed.
		\label{fig:phaseyear}
	}
\end{figure}

\begin{figure}[tp]
	\begin{center}
		\leavevmode
		\includegraphics[width=0.9\textwidth]{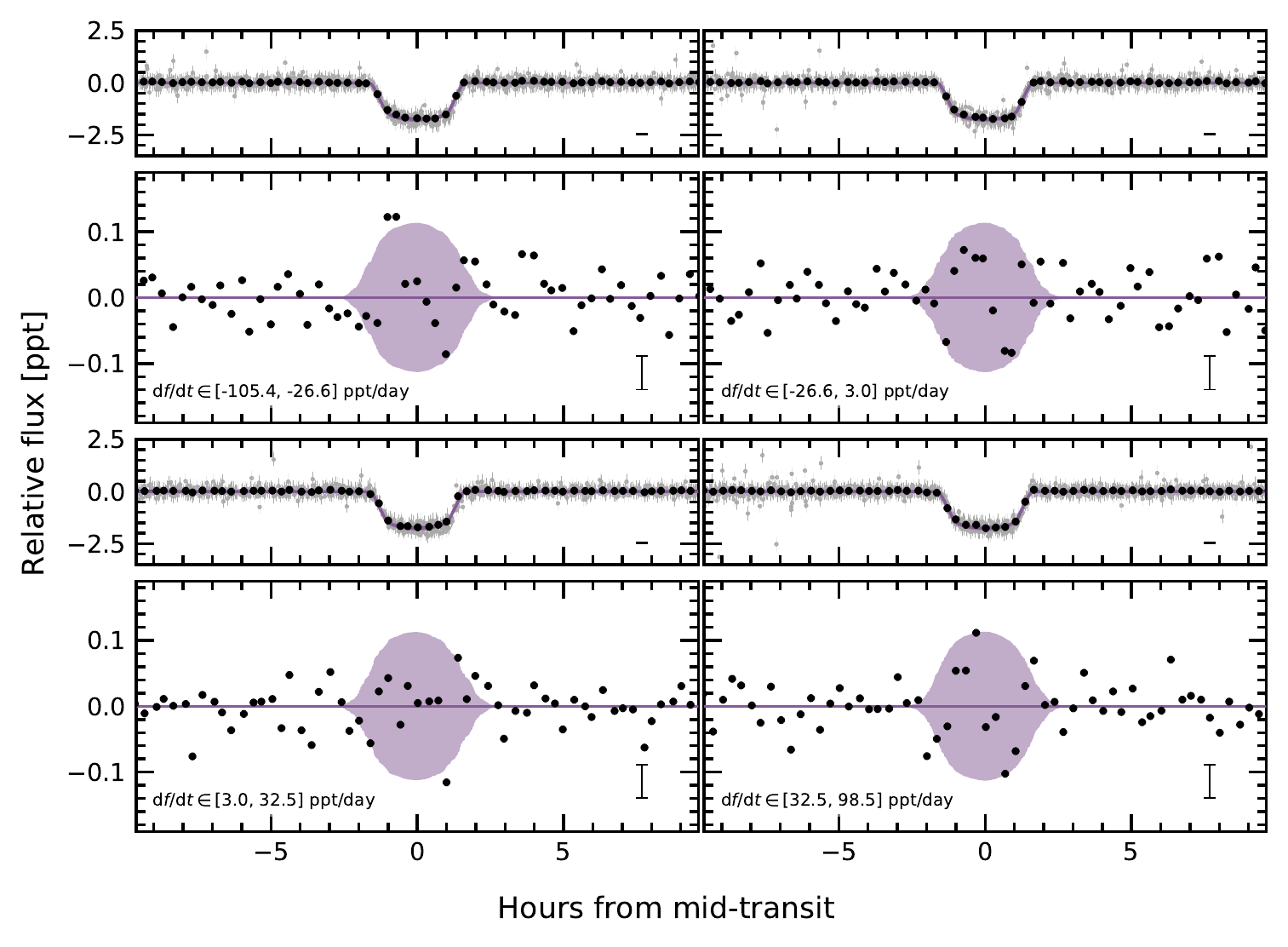}
	\end{center}
	\vspace{-0.7cm}
	\caption{
    {\bf Transit models and residuals, binned by quartiles in the
    local slope of the light curve}.  
    Representative uncertainties for the black points (binned at 20
    minute intervals) are shown in the lower right of each panel.  A
    similar transit asymmetry to that shown in Figure~\ref{fig:lc}
    seems to be present in three of the four bins.
		\label{fig:phaseslope}
	}
\end{figure}

\subsection{How Robust is the Asymmetric Transit?}

As a means of exploring the robustness of the transit asymmetry,
Figures~\ref{fig:phasequarter},~\ref{fig:phaseyear}, and~\ref{fig:phaseslope}
show the Kepler data binned in three ways: over
Kepler quarters, Julian years, and quartiles of local slope.  Over Kepler quarters
(Figure~\ref{fig:phasequarter}), Quarter 6 shows the strongest
asymmetry: a deviation of about 3\,ppt from expectation.  Quarter 7
shows an anomaly at roughly the same transit phase.  Year 2
correspondingly shows the strongest anomaly out of any year in
Figure~\ref{fig:phaseyear}; the asymmetry is visually apparent however
in each of the four years.

To bin by quartiles in local slope, we used our measurements
of the local linear slopes in each of the observed transit windows
(144 transits total).  Four outlier transits were removed, leaving 140
transits.  These were then divided into quartiles, so that each panel
shows 35 transits binned together.  The exact light curve slope intervals are listed in the lower left panels of
Figure~\ref{fig:phaseslope}.  Binned by local
slope quartiles (Figure~\ref{fig:phaseslope}), the asymmetry is
visually present in three of the four quartiles: the only bin in which
it does not appear is ${\rm d}f/{\rm d}t\in[3.0,32.5]\,{\rm
ppt\,day}^{-1}$.

Within the theory presented by \citet{mazeh_time_2015}, unresolved
starspot crossings cause the weak correlation between TTVs and the local
light curve slope (Figure~\ref{fig:corr}).  In this model, we would
expect the light curves with the most negative local slopes to have
the largest positive TTVs, due to spot crossing events during the
latter half of transit.  The upper-left panel of
Figure~\ref{fig:phaseslope} agrees with this expectation.  However, we
would also expect the sign of the effect to reverse when considering
the most positive local slopes (most negative TTVs).  The
lower-right panel of
Figure~\ref{fig:phaseslope} contradicts this expectation: the residual
in both cases maintains the same parity!  On the one hand, this shows
that the residual is not dependent on the local light curve slope,
which lowers the likelihood that it might be an artifact of our
detrending methods.  On the other, it raises the question of whether
unresolved starspot crossings are indeed the root cause of the
correlation shown in Figure~\ref{fig:corr}.
While we do not have a solution to this contradiction, 
the injection-recovery tests discussed in
Section~\ref{subsec:ttvslope} provide some assurance that the
TTV-slope correlation is not simply a systematic artifact.

\subsection{Interpretation}

The transit asymmetry seems robust against most methods of binning the
data, though with some caveats ({\it e.g.}, the ``middle quartile'' in
local flux, ${\rm d}f/{\rm d}t\in[3.0,32.5]\,{\rm ppt\,day}^{-1}$,
where the asymmetry does not appear).  Nonetheless, if the asymmetric were
systematic we might expect its parity to reverse as a function of
the sign of the local slope, and it does not.  We therefore entertained four
possible astrophysical explanations: gravity darkening, transit
timing variations, spot-crossing events, and a persistent asymmetric
dusty outflow.  

Gravity darkening is based on the premise that the rapidly rotating
star is oblate, and brighter near the poles than the equator
\citep[{\it e.g.},][]{masuda_spin-orbit_2015}.  The fractional transit
shape change due to gravity darkening is on the order of $(P_{\rm
break}/P_{\rm rot})^2$, for $P_{\rm break}$ the break-up rotation
period, and $P_{\rm rot}$ the rotation period.  Using the parameters
from Table~\ref{tab:posterior}, this yields an expected 0.14\%
distortion of the $\approx$1.8\,ppt transit depth: {\it i.e.}, an
absolute deviation of $\approx$2.5\,ppm.  The observed residual has a
semi-amplitude of $\approx 50\,{\rm ppm}$.  Since the expected signal
is smaller than the observed anomaly by over an order of magnitude,
gravity darkening seems to be an unlikely explanation.

The scenario of transit timing variations (TTVs) producing the
asymmetry seems unlikely because the transit
timing variations we do observe are correlated with the local light
curve slope, which increases roughly as much as it decreases.  From
our analysis, the mean TTV and its standard deviation are
$0.66\pm5.53$\,min; similarly the mean local slope and its
standard deviation are $0.59\pm45.50$\,ppt\,day$^{-1}$.  There is
therefore little expectation for TTVs to produce the asymmetry.  A
separate line of argument comes from Figure~\ref{fig:phaseslope}.  If
the local slope were essential to producing the transit asymmetry, we
would expect that in the largest ${\rm d}f/{\rm d}t$ bin, ${\rm
d}f/{\rm d}t\in[3.0,32.5]\,{\rm ppt\,day}^{-1}$, the sign of the
asymmetry would reverse.  We do not see evidence for this being the
case.

The third and related possibility is that of starspot crossings.
Young stars have higher spot-covering fractions than old stars ({\it
e.g.}, \citealt{morris_relationship_2020}).  Young solar-type stars
may also host dark starspots at high stellar latitudes \citep[{\it
e.g.}, EK~Dra;][]{strassmeier_starspots_2009}, though interferometric
imaging of spotted giant stars has shown different starspot latitude
distributions than those inferred from Doppler imaging
\citep{roettenbacher_contemporaneous_2017}.  Regardless, for any
spot-crossing anomalies to add coherently over the 140 Kepler
transits, it seems likely that we would need either for spots to be
persistent at a particular latitude (and for the planetary orbit to be
somewhat misaligned), or for a ``stroboscopic'' longitudinal phasing
\citep[{\it e.g.},][]{dai_stellar_2018}.  For our system, $P_{\rm
orb}/P_{\rm rot} \approx 2.76$, which means that every 4 transits and
11 stellar rotations, the planet crosses over roughly the same stellar
longitude, which might enable the necessary phasing if the spot-groups
are large and long-lived.  Unfortunately, the  S/N per Kepler transit
is $\approx8$, which renders individual spot-crossing events
unresolved.  This explanation seems marginally plausible, mainly
because the expected spot-crossing anomaly amplitudes ($\approx
100\,{\rm ppm}$) resemble the observed amplitude of the asymmetry
($\approx 50\,{\rm ppm}$).  One issue with this explanation however is
that there is no reason to expect starspot crossing events to last
exactly half the transit duration.

A persistent feature of the planet itself might therefore be needed to
explain the transit asymmetry.  An asymmetric outflow from the
planet's atmosphere could at least geometrically meet the requirements
\citep[{\it e.g.},][]{mccann_2019}.  To explain the asymmetric
transit, a small, dense component would lead the planet, and a long,
more rarefied (and variable) component would trail it.  This might
also explain the slight flux decrement visible for $\sim$1 hour
pre-ingress (Figure~\ref{fig:lc}).  The amplitude of the asymmetry is
roughly in line with theoretical expectations for dusty outflows
\citep{wang_dai_2019}, and based on the planet's size, its mass is
likely in a regime where such outflows are possible.  Out of the four
explanations discussed, this one at least theoretically seems the most
plausible.  By composition, the expectation would be that the envelope
is mostly hydrogen and helium gas, with a dust or haze component
providing the broadband opacity in the Kepler bandpass.  A natural
path for testing this idea would be to observe
additional transits of the planet in hydrogen absorption, metastable
helium absorption, or across a broad wavelength range in the
near-infrared.

%
%

\section{Spectroscopy and Photometry During the Transit of 2021 Aug 7}
\label{app:groundobs}

\begin{figure*}[tp]
	\begin{center}
		\leavevmode
		\subfloat{
			\includegraphics[width=\textwidth]{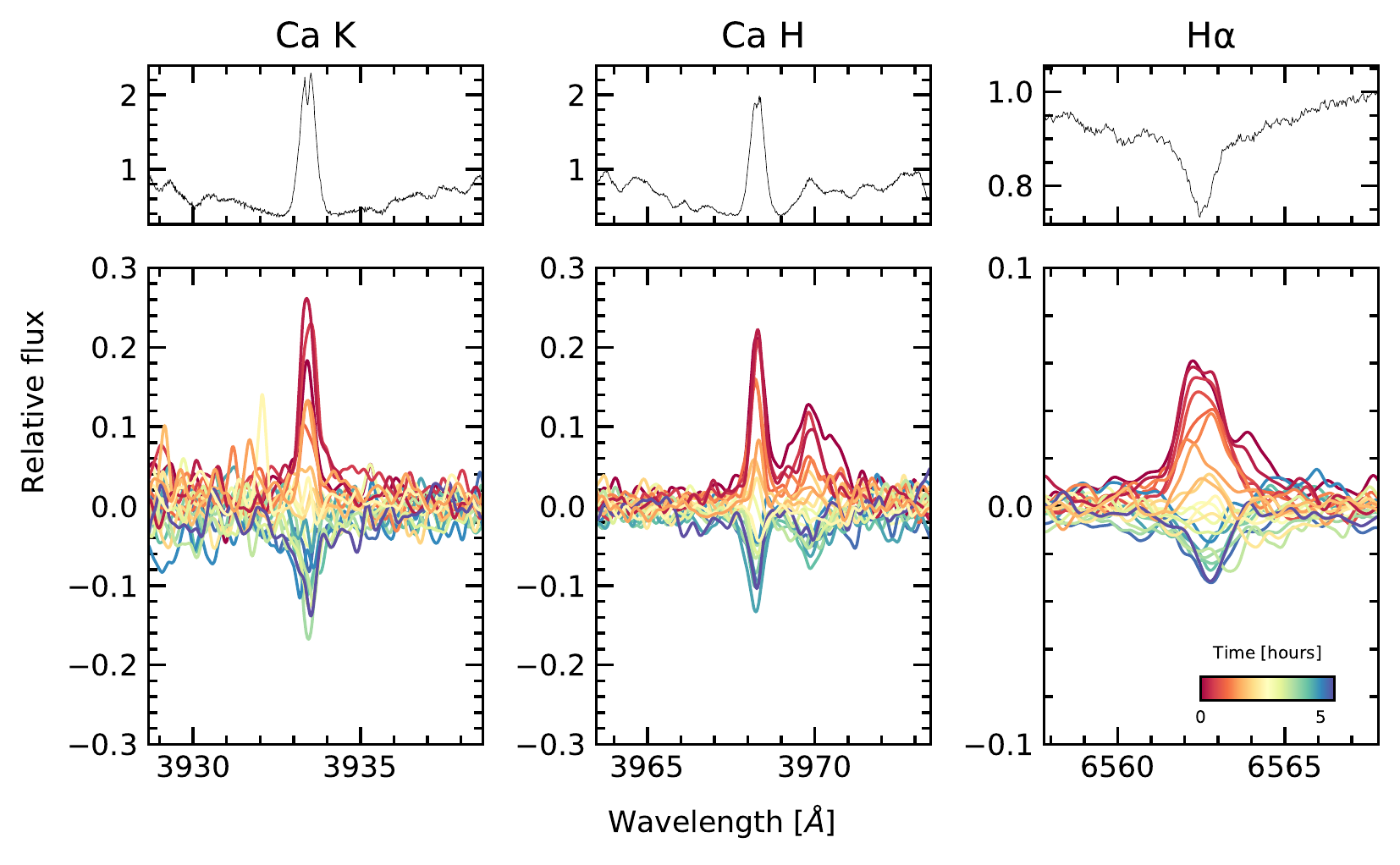}
		}
	\end{center}
	\vspace{-0.5cm}
	\caption{
    {\bf Spectroscopic activity indicators during the transit of 2021
    Aug 7.} The {\it top panels} show the median line profiles Ca K,
    Ca H, and H$\alpha$ line profiles from the HIRES spectra.  The
    {\it lower panels} show the differences of each individual
    spectrum relative to the median spectrum.  The bump in the red
    wing of Ca H is H$\epsilon$.  The spectra in the lower panels are
    smoothed for visualization purposes.
    \label{fig:rvactivity}
	}
\end{figure*}

\added{We used the ephemeris of \citet{holczer_transit_2016} to
observe a transit of \pn\ on the night of 2021 Aug 7 both
spectroscopically and photometrically.  We used the HIRES echelle
spectrograph at the Keck-I telescope and the MuSCAT3 photometer at Haleakal\=a
Observatory on Maui, HI \citep{Narita_2020}.} \deleted{ We monitored
\sn\ with Keck/HIRES before, during, and after transit on the night of
2021 Aug 7.} \replaced{We used the iodine cell for wavelength
calibration}{For the HIRES wavelength calibration, we used the iodine
cell}, and extracted the 1-D spectra using the standard California
Planet Survey pipeline \citep{howard_cps_2010}.  \added{Given the
faintness of the target ($V=13.1$), we observed using the C2 decker,
which yielded an instrument resolution of $\approx$50{,}000.} The
airmass ranged between 1.1 and 2.2 from the start through the end of
observations; the seeing ranged from $1\farcs1$ at the beginning to
$1\farcs5$ at the end.  \added{The HIRES exposure time was set at
$\approx$15 minutes in order to resolve the 2.8 hour transit event,
which yielded a S/N per resolution element of $\approx 75$ (low for
precision radial velocity standards).} \replaced{We also
simultaneously observed across {\it griz} bands using MuSCAT3 at
Haleakal\=a Observatory on Maui, HI.}{For the MuSCAT3 observations, we
observed simultaneously across the {\it griz} bands.} \added{The
exposure times in each bandpass ranged between 23 and 46 seconds, and
were chosen in order to yield a S/N in the peak pixel that exceeded
130 while also preventing saturation.} Performing aperture photometry
on the latter image stack yielded the data given in
Table~\ref{tab:phot}.

We considered two approaches to measuring the velocities: in the
first, hereafter ``Method 1'', we cross-correlated against a template
found via spectral classification with \texttt{SpecMatch-Emp}
\citep{yee_SM_2017}.  In ``Method 2'', we used a high S/N template of
V1298~Tau.  Although V1298 Tau is cooler than Kepler~1627A by
$\approx$500\,K, it has a comparable amount of line-broadening ($v\sin
i = 23$\kms), and a comparable level of stellar activity.  The mean
and standard deviation of the internal RV uncertainties averaged over
all epochs were $16.2\pm1.1$\ms\ from Method 1, and $12.6\pm0.6$\ms\
from Method 2.  The corresponding time-averaged reduced $\chi^2$ from
the template match was $1.57\pm0.04$ (Method 1) and $1.30\pm0.02$
(Method 2).  Given these diagnostics, we adopted the velocities from
the second approach, which are reported in Table~\ref{tab:rv}.

Figure~\ref{fig:ground} shows the results.  The MuSCAT3 photometry
shows the expected starspot trend, along with the transit and what is
likely a chromatic starspot crossing event at ${\rm JD} - 2459433 =
0.955$.  The radial velocities decrease by $\approx$250\ms\ over the
six hour window.  This decrease in RV is correlated with a decrease in
the S-indices derived from the Ca HK lines.  One outlying RV point is
apparent shortly before egress; it is temporally coincident with an
outlying value in the S-index time series.

Overall, we expect the dominant trends in both the photometry and
radial velocities to be caused by starspots on the stellar photosphere
rotating into and out of view.  The plasma in the leading and receding
limbs of the stellar disk has an apparent line-of-sight velocity of
$\pm 20$\kms.  Over 10\% of a rotation cycle ($P_{\rm rot}=2.6\,{\rm
days}$), spots near these limbs come into and out of view, modulate
the stellar velocity profile, and can thereby produce the overall
$\approx$250\ms\ trend.  The Ca HK and H$\alpha$ emission profiles
support this interpretation; Figure~\ref{fig:rvactivity} shows that
each line gradually decreases in intensity over the course of the six
hour sequence.

The expectation however is for the starspot-induced signals to be
smooth, at worst with contributions at $0.5\,P_{\rm rot}$ or
$0.25\,P_{\rm rot}$ \citep{klein_simulated_2020}.  We therefore fitted
the RVs using the \citet{hirano_analytic_2010,hirano_2011} models for
the Rossiter-McLaughlin (RM) effect, and allowed for an optional
linear and quadratic trend in time to fit the $\approx$250\ms\
spot-induced trend.  We followed the methodology developed by
\citet{stefansson_2020}.  We allowed the sky-projected obliquity, the
projected stellar equatorial velocity, and the Gaussian dispersion of
the spectral lines to vary, and fixed the limb-darkening using the
$V$-band tabulation from \citet{claret_gravity_2011}.  We assumed a
Gaussian prior on $v\sin i$ and $a/R_\star$ from
Table~\ref{tab:starparams}, and also allowed for a white-noise jitter
term to be added in quadrature to the measurement uncertainties.  We
used a 15\,minute exposure time to numerically evaluate the model.

The quadratic model with the RM effect is shown in
Figure~\ref{fig:ground}; the jitter term is incorporated in the model
uncertainties, but not the plotted measurement uncertainties.  The
plotted measurement uncertainties are the internal uncertainties on
the RVs ($\approx 13\,{\rm m\,s}^{-1}$), and are dominated by the
$v\sin i$ broadening.  However, between exposures, the RVs show
significant additional scatter that is not captured by the slow
quadratic trend.  The white-noise jitter for this particular model is
$\sigma_{\rm RV} = 27^{+6}_{-5}$\ms, which is comparable to the
expected RM anomaly of $\Delta v_{\rm RM} \approx f_{\rm LD} \cdot
\delta \cdot v\sin i \cdot \sqrt{1-b^2} \approx 20\,{\rm m\,s}^{-1}$,
assuming a perfectly aligned orbit.

The presence of this additional scatter prevents a convincing
detection of the RM effect.  The reason can be understood via model
comparison.  If we compare the model with a quadratic trend and the RM
effect against a model with a linear trend and the RM effect, or even
a model with no RM effect at all, then the respective Bayesian
Information Criterion (BIC) values are as follows.
\begin{align}
  {\rm BIC} &= 227.1  \quad ({\rm Quadratic+RM}) \nonumber \\
  {\rm BIC} &= 231.1  \quad ({\rm Linear+RM}) \nonumber \\
  {\rm BIC} &= 221.4  \quad ({\rm Only\ Quadratic}).
\end{align}
There is therefore no evidence to prefer the model with the RM effect
against a model that only accounts for the stellar variability.  The
``only quadratic'' model does particularly well because it can inflate
the jitter term to account for scatter during the transit (even if the
scatter contains astrophysics!), and it has fewer free parameters.
However, we cannot justify a physical prior on the jitter term,
because we do not understand the origin of the exposure-to-exposure
scatter.  As noted above, the velocity deviations from starspots are
expected to have contributions at the stellar rotation frequency, or
harmonics thereof.  This jitter is present on the exposure timescale
(15 minutes), which is only 0.4\% of the stellar rotation period; it
is not obvious that starspots would be the culprit.

The amplitude of both the spot-induced trend and the jitter are
somewhat larger than recent comparable measurements in systems such as
AU~Mic \citep{palle_transmission_2020}, DS~Tuc
\citep{montet_young_2020,zhou_well_2020} and TOI~942
\citep{wirth_2021_toi942}.  One possible explanation for the jitter is
that it is astrophysical in origin, and that it is caused by some
novel process operating on the surface of Kepler 1627A.  Another
possibility is that our RV analysis underestimates our measurement
uncertainties; in order to achieve the requisite time-sampling the S/N
per resolution element in our spectra was 70 to 80, which is lower
than desired for deriving high-precision velocities.  In addition, the
rapid rotation of the star could affect accuracy of the uncertainties
from the velocity extraction.  Observations at higher S/N are
necessary to distinguish these two possibilities, and remain
worthwhile in order to clarify the orbital geometry of \pn.  Useful
next steps would include transit observations with a stabilized
spectrograph in the optical \citep[{\it
e.g.},][]{gibson_2016_kpf,seifahrt_maroon-x_2018}, or in the
near-infrared \citep[{\it e.g.},][]{feinstein_halpha_2021}.

\input{phot_table.tex}
\input{rv_table.tex}

\section{Flare Analysis}
\label{app:flare}

\begin{figure*}[tp]
	\begin{center}
		\leavevmode
		\includegraphics[width=0.9\textwidth]{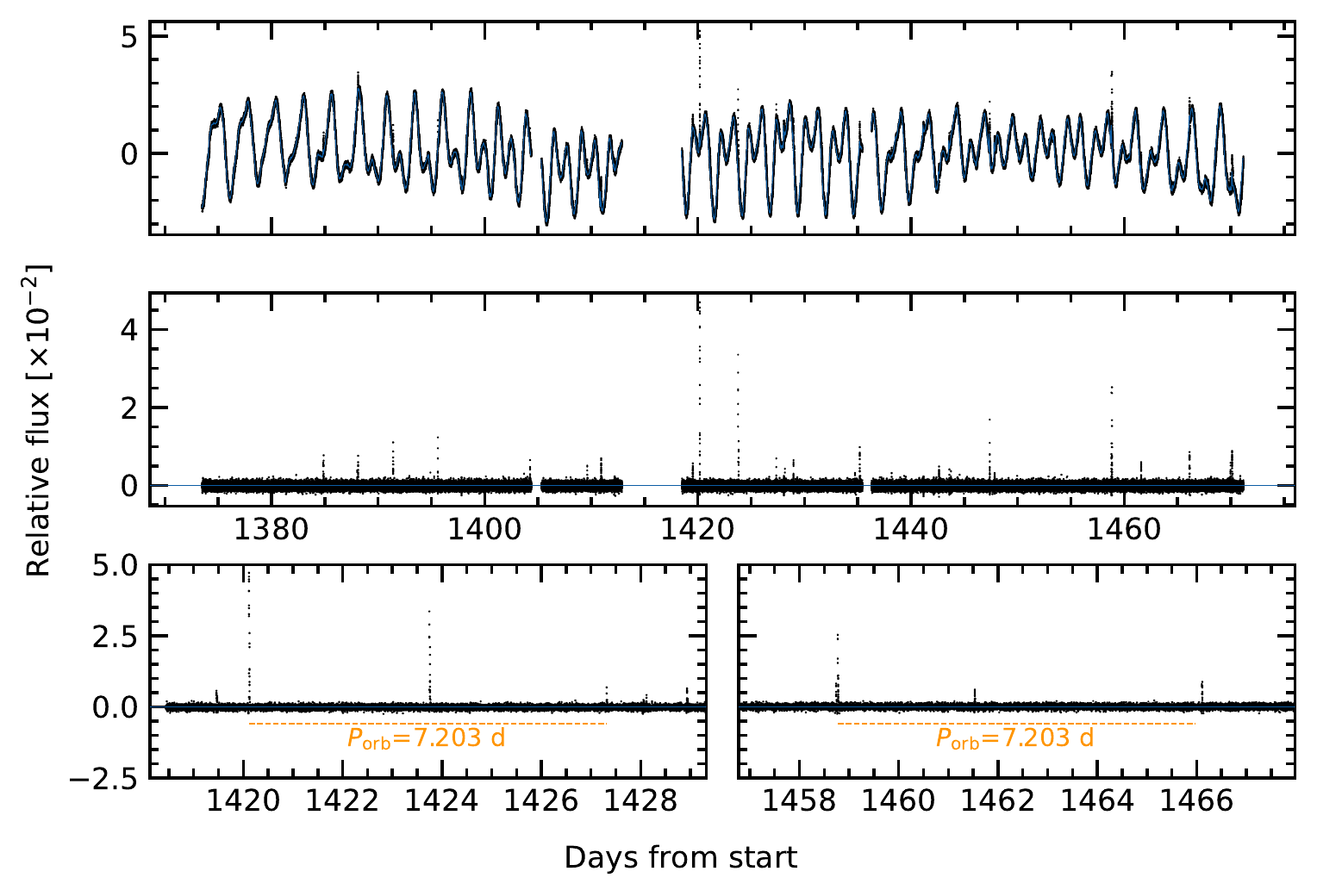}
	\end{center}
	\vspace{-0.7cm}
	\caption{
		{\bf Flares in Kepler\,1627}.  
		{\it Top:}
		The 1-minute cadence Kepler data (black points) is shown with a stellar variability model superposed
		(blue line).
		{\it Middle:}
		Residual after subtracting the stellar variability model.  Flares
		appear as spikes.
		{\it Bottom:}
		Zooms of the brightest, and third-brightest flares.  A timing
		coincidence -- that both flares have ``successors'' approximately
		one \added{planetary }orbital period after the initial event -- is emphasized.
		\label{fig:flarezoom}
	}
\end{figure*}


\begin{figure*}[tp]
	\begin{center}
		\leavevmode
		\includegraphics[width=0.8\textwidth]{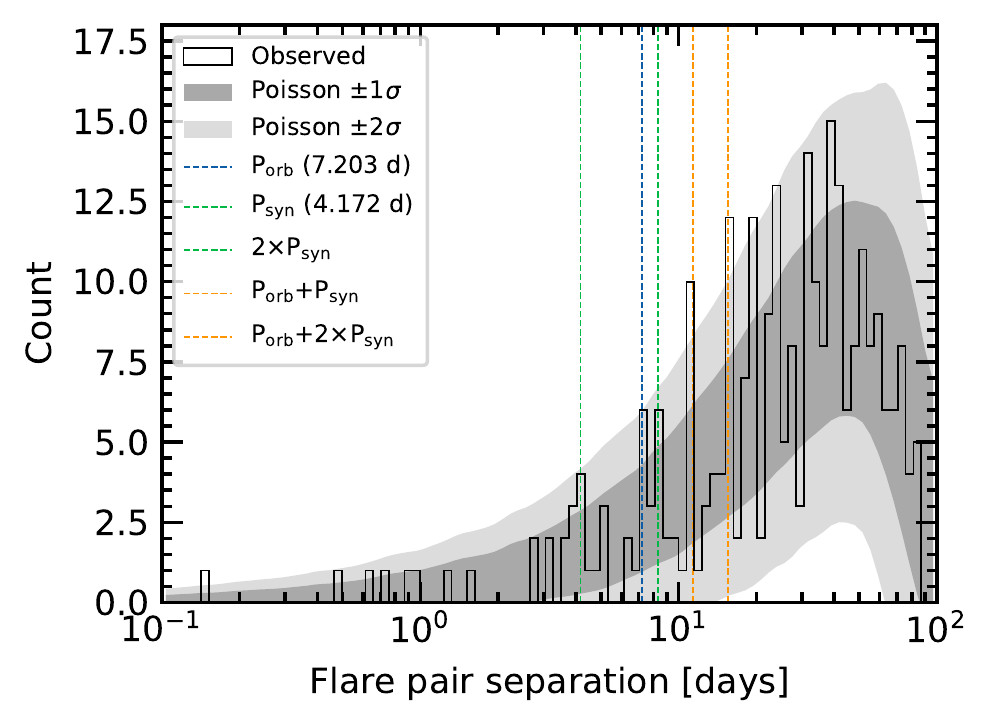}
	\end{center}
	\vspace{-0.7cm}
	\caption{
		{\bf Statistics of inter-flare arrival times}.  
		24 flares were recorded with amplitudes exceeding 0.5\% over the
		97.7 days of short cadence observations.  The histogram of the
		time intervals between every possible pair of flares is shown in
		black.  Some plausibly important timescales for star-planet
		interactions, namely the planetary orbital period and synodic
		period (the orbital period as seen from the rotating stellar
		frame) are shown along with their linear combinations.  Monte
		Carlo draws from a Poisson distribution are shown with the gray
		bands.  While peaks in the observed distribution do coincide with
		some of the ``special periods'', the statistical evidence for a
		non-Poissonian process driving the flares does not clear the
		$5$-$\sigma$ threshold.
		\label{fig:flarestats}
	}
\end{figure*}

In addition to the 3.9 years of long cadence data, short cadence
(1-minute) Kepler observations were acquired over 97.7 days during
Quarter 15.  The short cadence light curve shows a higher rate of
flaring than visible in the long cadence data
(Figure~\ref{fig:flarezoom}).  We analyzed the short cadence light
curve and its flares according to the following procedure.
\begin{enumerate}
  \item Fit the starspot-induced variability using a Gaussian Process
    with a \texttt{SHOTerm} kernel, a white-noise jitter term, and the
    mean flux.
  \item Select points more than twice the median absolute
    deviation from the residual, and exclude them from the light
    curve (these points include the flares).  Repeat Step~1.
  \item Using the residual from Step 2, identify all flares, requiring
    them to be at least 20 cadences apart, at least 7 median absolute
    deviations above the median baseline, and lasting at least 2
    cadences in duration.  Build the mask spanning these times, from 5
    minutes before each flare begins to 2.5 minutes after the final
    flare cadence.  Repeat Step 1 a final time.
\end{enumerate}
The final step of flare identification and fitting was performed using
\texttt{altaipony} \citep{davenport_2016,ilin_flares_2021}.  The
analytic flare model is from \citet{davenport_2014} and it
parametrizes the flare with a start time, an exponential lag time, and
an amplitude.

There were $N_{\rm f}=24$ flares that exceeded $0.5\%$ in relative
flux during the short cadence observations.  These 24 flares spanned a
total of 6.5 hours ($\sim$15 minutes per flare).  Inspecting the data,
we noticed a coincidence in the flare arrival times.  The coincidence
is that despite the low flare duty cycle, one orbital period after the
brightest flare, a second flare followed.  This and a similar event
are shown in Figure~\ref{fig:flarezoom}.  The timing error is good to
a $\approx0.2\%$ difference from the orbital period, which given the
duty cycle seems {\it a priori} unlikely.  If we consider flares
falling within 2\% of the planet's orbital period after a previous
flare, then 4 of the 24 flare events have candidate ``successors''.

As with any coincidence, if one does not have a firm prediction, it is
difficult to assess whether a surprise is statistically significant.
Since our surprise was specifically at the inter-arrival time of
certain flares coinciding with special time intervals, we performed
the following analysis.  First, we considered all unordered pairs of
flares.  For $N$ flares there are ${n \choose 2}$ such pairs (for our
case, 276 pairs).  We then compared the distribution of the pair
separations against that of a Poisson distribution.  Specifically, we
drew $N_{\rm f}=24$ samples from a Poisson distribution with $\lambda
= \Delta t / N_{\rm f}$, for $\Delta t=97.7\,{\rm days}$ the total
duration of the observations, and repeated the draw $10^3$ times with
unique random seeds.

Figure~\ref{fig:flarestats} shows the results.  The vertical lines in
the figure show the planetary orbital period, the synodic period
$P_{\rm syn} = (P_{\rm rot}^{-1} - P_{\rm orb}^{-1})^{-1}$, and linear
combinations thereof.  The tidal period (half the synodic period) is
not shown.  The bins are logarithmically spaced to give 100 bins
between the minimum and maximum ordinate values.  The gray bands
express the range of values observed from the Poissonian draws.  While
it does seem like an odd coincidence for peaks in the observed flare
arrival time distribution to coincide with the locations of these
``special intervals'', the statistical evidence for a non-Poissonian
process driving the flares does not seem especially overwhelming.
More quantitatively, the peaks observed at the orbital and synodic
periods are within the $\pm 2$-$\sigma$ range of a Poissonian process,
and those at $P_{\rm orb}+P_{\rm syn}$ and $P_{\rm orb}+2P_{\rm syn}$
are only slightly above this range.  With that said, future analyses
of these data by investigators with more knowledge of this topic could
very well yield more quantitative insights.  Such analyses should keep
in mind an important caveat: the amplitude distribution of M-dwarf
flares extends up to many times the quiescent flux \citep[see Figure~7
of][]{gunther_2020}.  A flare on Kepler\,1627B producing double its
quiescent white-light flux would yield a $\approx$1\% apparent
amplitude.  Such flares could represent a significant fraction of
those in the Kepler observations.


\end{document}

%% file: authors.tex
\correspondingauthor{L.\,G.\,Bouma}
\email{luke@astro.caltech.edu}

\author[0000-0002-0514-5538]{L. G. Bouma}
\altaffiliation{51 Pegasi b Fellow}
\affiliation{Cahill Center for Astrophysics, California Institute of Technology, Pasadena, CA 91125, USA}
\affiliation{Department of Astrophysical Sciences, Princeton University, 4 Ivy Lane, Princeton, NJ 08540, USA}

\author[0000-0002-2792-134X]{J. L. Curtis}
\affiliation{Department of Astronomy, Columbia University, 550 West 120th Street, New York, NY 10027, USA}
\affiliation{Department of Astrophysics, American Museum of Natural History, New York, NY 10024, USA}

\author[0000-0003-1298-9699]{K. Masuda}
\affiliation{Department of Earth and Space Science, Osaka University, Osaka 560-0043, Japan}

\author{L. A. Hillenbrand}
\affiliation{Cahill Center for Astrophysics, California Institute of Technology, Pasadena, CA 91125, USA}

\author[0000-0001-7409-5688]{G. Stefansson}
\altaffiliation{Henry Norris Russell Fellow}
\affiliation{Department of Astrophysical Sciences, Princeton University, 4 Ivy Lane, Princeton, NJ 08540, USA}

%
%
%
\author[0000-0002-0531-1073]{H. Isaacson}
\affiliation{Astronomy Department, University of California, Berkeley, CA 94720, USA}

%
%
\author[0000-0001-8511-2981]{N. Narita}
\affiliation{Komaba Institute for Science, The University of Tokyo, Tokyo 153-8902, Japan}
\affiliation{Japan Science and Technology Agency, PRESTO, Tokyo 153-8902, Japan}
\affiliation{Astrobiology Center, Tokyo 181-8588, Japan}
\affiliation{Instituto de Astrof\'{i}sica de Canarias (IAC), 38205 La Laguna, Tenerife, Spain}

\author[0000-0002-4909-5763]{A. Fukui} 
\affiliation{Komaba Institute for Science, The University of Tokyo, Tokyo 153-8902, Japan}
\affiliation{Instituto de Astrof\'{i}sica de Canarias (IAC), 38205 La Laguna, Tenerife, Spain}

\author[0000-0002-5658-5971]{M. Ikoma} 
\affiliation{Division of Science, National Astronomical Observatory of Japan, Tokyo 181-8588, Japan}

\author[0000-0002-6510-0681]{M. Tamura} 
\affiliation{Department of Astronomy, The University of Tokyo, Tokyo 113-0033, Japan}
\affiliation{Astrobiology Center, Tokyo 181-8588, Japan}
\affiliation{Division of Science, National Astronomical Observatory of Japan, Tokyo 181-8588, Japan}

\author[0000-0001-9811-568X]{A.~L.~Kraus}
\affiliation{Department of Astronomy, The University of Texas at Austin, Austin, TX 78712, USA}

\author[0000-0001-9800-6248]{E. Furlan} 
\affiliation{NASA Exoplanet Science Institute, Caltech/IPAC, Pasadena, CA 91125, USA}

\author[0000-0003-2519-6161]{C.~L.~Gnilka} 
\affiliation{NASA Ames Research Center, Moffett Field, CA 94035, USA}
\affiliation{NASA Exoplanet Science Institute, Caltech/IPAC, Pasadena, CA 91125, USA}

\author[0000-0002-9903-9911]{K.~V.~Lester} 
\affiliation{NASA Ames Research Center, Moffett Field, CA 94035, USA}

\author[0000-0002-2532-2853]{S. B. Howell}
\affiliation{NASA Ames Research Center, Moffett Field, CA 94035, USA}

%% file: starparams.tex
\begin{table*}
\scriptsize
\setlength{\tabcolsep}{2pt}
\centering
\caption{Literature and Measured Properties for Kepler$\,$1627}
\label{tab:starparams}
\begin{tabular}{llcc}
  \hline
  \hline
Primary Star\dotfill & \\
\multicolumn{3}{c}{TIC 120105470} \\
\multicolumn{3}{c}{GAIADR2$^\dagger$ 2103737241426734336} \\
\hline
\hline
Parameter & Description & Value & Source\\
\hline 
$\alpha_{J2015.5}$\dotfill	&Right Ascension (hh:mm:ss)\dotfill & 18:56:13.6 & 1	\\
$\delta_{J2015.5}$\dotfill	&Declination (dd:mm:ss)\dotfill & +41:34:36.22 & 1	\\
V\dotfill			&Johnson V mag.\dotfill & 13.11 $\pm$ 0.08		& 2	\\
${\rm G}$\dotfill     & Gaia $G$ mag.\dotfill     & 13.02$\pm$0.02 & 1\\
$G_{\rm BP}$\dotfill     & Gaia $BP$ mag.\dotfill     & 13.43$\pm$0.02 & 1\\
$G_{\rm RP}$\dotfill     & Gaia $RP$ mag.\dotfill     & 12.44$\pm$0.02 & 1\\
${\rm T}$\dotfill     & TESS $T$ mag.\dotfill     & 12.53$\pm$0.02 & 2\\
J\dotfill			& 2MASS J mag.\dotfill & 11.69  $\pm$ 0.02	& 3	\\
H\dotfill			& 2MASS H mag.\dotfill & 11.30 $\pm$ 0.02	    &  3	\\
K$_{\rm S}$\dotfill			& 2MASS ${\rm K_S}$ mag.\dotfill & 11.19 $\pm$ 0.02 &  3	\\
$\pi$\dotfill & Gaia EDR3 parallax (mas) \dotfill & 3.009 $\pm$ 0.032 &  1 \\
$d$\dotfill & Distance (pc)\dotfill & $329.5 \pm 3.5$ & 1, 4 \\
$\mu_{\alpha}$\dotfill		& Gaia EDR3 proper motion\dotfill & 1.716 $\pm$ 0.034 & 1 \\
                    & \hspace{3pt} in RA (mas yr$^{-1}$)	&  \\
$\mu_{\delta}$\dotfill		& Gaia EDR3 proper motion\dotfill 	&  -1.315 $\pm$ 0.034 &  1 \\
                    & \hspace{3pt} in DEC (mas yr$^{-1}$) &  \\
RUWE\dotfill		& Gaia EDR3 renormalized\dotfill 	&  2.899 &  1 \\
                    & \hspace{3pt} unit weight error &  \\
RV\dotfill & Systemic radial \hspace{9pt}\dotfill  & $-16.7 \pm 1.0$ & 5 \\
                    & \hspace{3pt} velocity (\kms)  & \\
Spec. Type\dotfill & Spectral Type\dotfill & 	G8V & 5 \\
$v\sin{i_\star}$\dotfill &  Rotational velocity$^*$ (\kms) \hspace{9pt}\dotfill &  18.9 $\pm$ 1.0 & 5 \\
Li EW\dotfill & 6708\AA\ Equiv{.} Width (m\AA) \dotfill & $233^{+5}_{-7}$  & 5 \\
$T_{\rm eff}$\dotfill &  Effective Temperature (K) \hspace{9pt}\dotfill & 5505 $\pm$ 60 &  6  \\
$\log{g_{\star}}$\dotfill &  Surface Gravity (cgs)\hspace{9pt}\dotfill &  4.53 $\pm$ 0.05  &  6 \\
$R_\star$\dotfill & Stellar radius ($R_\odot$)\dotfill & 0.881$\pm$0.018 & 6 \\
$M_\star$\dotfill & Stellar mass ($R_\odot$)\dotfill & 0.953$\pm$0.019 & 6 \\
$A_{\rm V}$\dotfill & Interstellar reddening (mag)\dotfill & 0.2 $\pm$ 0.1 & 6 \\
${\rm [Fe/H]}$\dotfill &   Metallicity\dotfill & 0.1 $\pm$ 0.1 & 6 \\
$P_{\rm rot}$\dotfill & Rotation period (d)\dotfill & $2.642\pm 0.042$  & 7 \\
Age & Adopted stellar age (Myr)\dotfill & $38^{+6}_{-5}$  &  8 \\
\hline 
$\Delta m_{832}$ & Mag difference (`Alopeke 832\,nm)\dotfill & $3.14 \pm 0.15$ & 9 \\
$\theta_{\rm B}$ & Position angle (deg)\dotfill & $92 \pm 1$ & 9 \\
$\rho_{\rm B}$ & Apparent separation of \dotfill & $0.164 \pm 0.002$ &  9 \\
                    & \hspace{3pt} primary and secondary (as) &  \\
$\rho_{\rm B}$ & Apparent separation of \dotfill & $53 \pm 4$ &  1,4,9 \\
                    & \hspace{3pt} primary and secondary (AU) &  \\
$\Delta m_{K'}$ & Mag difference (NIRC2 $K'$)\dotfill & $2.37 \pm 0.02$ & 10 \\
$\theta_{\rm B}$ & Position angle (deg)\dotfill & $95.9 \pm 0.5$ & 10 \\
$\rho_{\rm B}$ & Apparent separation of \dotfill & $0.1739 \pm 0.0010$ &  10 \\
                    & \hspace{3pt} primary and secondary (as) &  \\
\hline
\end{tabular}
\begin{flushleft}
 \footnotesize{ \textsc{NOTE}---
 $^\dagger$ The GAIADR2 and GAIAEDR3 identifiers for Kepler 1627A are identical.  The secondary
 is not resolved in the Gaia point source catalog.
 $^*$ Given only $v\sin i$ and $2\pi R_\star/P_{\rm rot}$, $\cos i=0.11^{+0.11}_{-0.08}$.
Provenances are:
$^1$\citet{gaia_collaboration_2021_edr3},
$^2$\citet{stassun_TIC8_2019},
$^3$\citet{skrutskie_tmass_2006},
$^4$\citet{Lindegren_2021_offset},
$^5$HIRES spectra and \citet{yee_SM_2017},
$^6$Cluster isochrone (MIST adopted; PARSEC compared for quoted
  uncertainty),
$^7$Kepler light curve,
$^8$Pre-main-sequence CAMD interpolation (Section~\ref{sec:camd}),
$^9$`Alopeke imaging 2021 June 24 \citep{scott_twin_2021},
$^{10}$NIRC2 imaging 2015 July 22, using the \citet{yelda_2010} optical distortion solution to convert pixel-space relative positions to on-sky relative astrometry.
The ``discrepancy'' between the two imaging epochs likely indicates orbital motion.

}
\end{flushleft}
\vspace{-0.5cm}
\end{table*}

%% file: model_posterior_table.tex
\begin{deluxetable*}{lllrrrrrrr}
  \tablecaption{ Priors and Posteriors for Model Fitted to the Long
  Cadence Kepler 1627Ab Light Curve.}
\label{tab:posterior}
\tabletypesize{\scriptsize}
%
%
\tablehead{
  \colhead{Param.} & 
  \colhead{Unit} &
  \colhead{Prior} & 
  \colhead{Median} & 
  \colhead{Mean} & 
  \colhead{Std{.} Dev.} &
  \colhead{3\%} &
  \colhead{97\%} &
  \colhead{ESS} &
  \colhead{$\hat{R}-1$}
}

\startdata
{\it Sampled} & & & & & & & & & \\
\hline
$P$ & d & $\mathcal{N}(7.20281; 0.01000)$ & 7.2028038 & 7.2028038 & 0.0000073 & 7.2027895 & 7.2028168 & 7464 & 3.9e-04 \\
$t_0^{(1)}$ & d & $\mathcal{N}(120.79053; 0.02000)$ & 120.7904317 & 120.7904254 & 0.0009570 & 120.7886377 & 120.7921911 & 3880 & 2.0e-03 \\
$\log \delta$ & -- & $\mathcal{N}(-6.3200; 2.0000)$ & -6.3430 & -6.3434 & 0.0354 & -6.4094 & -6.2767 & 6457 & 3.0e-04 \\
$b^{(2)}$ & -- & $\mathcal{U}(0.000; 1.000)$ & 0.4669 & 0.4442 & 0.2025 & 0.0662 & 0.8133 & 1154 & 1.6e-03 \\
$u_1$ & -- & \citet{exoplanet:kipping13} & 0.271 & 0.294 & 0.190 & 0.000 & 0.628 & 3604 & 1.5e-03 \\
$u_2$ & -- & \citet{exoplanet:kipping13} & 0.414 & 0.377 & 0.326 & -0.240 & 0.902 & 3209 & 1.4e-03 \\
$R_\star$ & $R_\odot$ & $\mathcal{N}(0.881; 0.018)$ & 0.881 & 0.881 & 0.018 & 0.847 & 0.915 & 8977 & 3.1e-04 \\
$\log g$ & cgs & $\mathcal{N}(4.530; 0.050)$ & 4.532 & 4.533 & 0.051 & 4.435 & 4.627 & 6844 & 1.6e-03 \\
$\langle f \rangle$ & -- & $\mathcal{N}(0.000; 0.100)$ & -0.0003 & -0.0003 & 0.0001 & -0.0005 & -0.0000 & 8328 & 1.1e-03 \\
$e^{(3)}$ & -- & \citet{vaneylen19} & 0.154 & 0.186 & 0.152 & 0.000 & 0.459 & 1867 & 2.0e-03 \\
$\omega$ & rad & $\mathcal{U}(0.000; 6.283)$ & 0.055 & 0.029 & 1.845 & -3.139 & 2.850 & 3557 & 8.6e-05 \\
$\log \sigma_f$ & -- & $\mathcal{N}(\log\langle \sigma_f \rangle; 2.000)$ & -8.035 & -8.035 & 0.008 & -8.049 & -8.021 & 9590 & 3.9e-04 \\
$\sigma_{\mathrm{rot}}$ & d$^{-1}$ & $\mathrm{InvGamma}(1.000; 5.000)$ & 0.070 & 0.070 & 0.001 & 0.068 & 0.072 & 9419 & 1.4e-03 \\
$\log P_{\mathrm{rot}}$ & $\log (\mathrm{d})$ & $\mathcal{N}(0.958; 0.020)$ & 0.978 & 0.978 & 0.001 & 0.975 & 0.980 & 8320 & 2.2e-04 \\
$\log Q_0$ & -- & $\mathcal{N}(0.000; 2.000)$ & -0.327 & -0.326 & 0.043 & -0.407 & -0.246 & 9659 & 2.7e-04 \\
$\log \mathrm{d}Q$ & -- & $\mathcal{N}(0.000; 2.000)$ & 7.697 & 7.698 & 0.103 & 7.511 & 7.899 & 5824 & 3.7e-04 \\
$f$ & -- & $\mathcal{U}(0.010; 1.000)$ & 0.01006 & 0.01009 & 0.00009 & 0.01000 & 0.01025 & 4645 & 4.0e-04 \\
\hline
{\it Derived} & & & & & & & & & \\
\hline
$\delta$ & -- & -- & 0.001759 & 0.001759 & 0.000062 & 0.001641 & 0.001875 & 6457 & 3.0e-04 \\
$R_{\rm p}/R_\star$ & -- & -- & 0.039 & 0.039 & 0.001 & 0.037 & 0.042 & 1811 & 1.1e-03 \\
$\rho_\star$ & g$\ $cm$^{-3}$ & -- & 1.990 & 2.004 & 0.240 & 1.570 & 2.461 & 6905 & 2.1e-03 \\
$R_{\rm p}^{(4)}$ & $R_{\mathrm{Jup}}$ & -- & 0.337 & 0.338 & 0.014 & 0.314 & 0.367 & 2311 & 1.0e-03 \\
$R_{\rm p}^{(4)}$ & $R_{\mathrm{Earth}}$ & -- & 3.777 & 3.789 & 0.157 & 3.52 & 4.114 & 2311 & 1.0e-03 \\
$a/R_\star$ & -- & -- & 17.606 & 17.619 & 0.702 & 16.277 & 18.906 & 6905 & 2.1e-03 \\
$\cos i$ & -- & -- & 0.027 & 0.025 & 0.010 & 0.004 & 0.040 & 1312 & 1.2e-03 \\
$T_{14}$ & hr & -- & 2.841 & 2.843 & 0.060 & 2.734 & 2.958 & 3199 & 3.6e-04 \\
$T_{13}$ & hr & -- & 2.555 & 2.539 & 0.094 & 2.360 & 2.692 & 1960 & 1.4e-03 \\
\enddata
\tablecomments{
  ESS refers to the number of effective samples.
  $\hat{R}$ is the Gelman-Rubin convergence diagnostic.
  Logarithms in this table are base-$e$.
  $\mathcal{U}$ denotes a uniform distribution,
  and $\mathcal{N}$ a normal distribution.
  (1) The ephemeris is in units of BJDTDB - 2454833.
  (2) Although $\mathcal{U}(0,1+R_{\rm p}/R_\star)$ is formally
  correct, for this model we assumed a non-grazing transit to enable
  sampling in $\log \delta$.
  (3) The eccentricity vectors are sampled in the $(e\cos\omega,
  e\sin\omega)$ plane.
  (4) The true planet size is a factor of $((F_1+F_2)/F_1)^{1/2}$
  larger than that from the fit because of dilution from Kepler
  1627B, where $F_1$ is the flux from the primary, and $F_2$ is that
  from the secondary; the mean and standard deviation of $R_{\rm
  p}=3.817\pm0.158\,R_{\oplus}$ quoted in the text includes this correction,
  assuming $(F_1+F_2)/F_1\approx 1.015$.
}
\vspace{-0.3cm}
\end{deluxetable*}

%% file: t1_appendix.tex
\begin{deluxetable*}{lll}
    

\tabletypesize{\scriptsize}


\tablecaption{Young, Age-dated, and Age-dateable Stars Within the
  Nearest Few Kiloparsecs.}
\label{tab:v06}


\tablehead{
  \colhead{Parameter} &
  \colhead{Example Value} &
  \colhead{Description}
}

%
\startdata
          \texttt{source\_id} &                                          1709456705329541504 &                                              Gaia DR2 source identifier. \\
                  \texttt{ra} &                                                      247.826 &                                          Gaia DR2 right ascension [deg]. \\
                 \texttt{dec} &                                                       79.789 &                                              Gaia DR2 declination [deg]. \\
            \texttt{parallax} &                                                       35.345 &                                                 Gaia DR2 parallax [mas]. \\
     \texttt{parallax\_error} &                                                        0.028 &                                     Gaia DR2 parallax uncertainty [mas]. \\
                \texttt{pmra} &                                                       94.884 &      Gaia DR2 proper motion $\mu_\alpha \cos \delta$ [mas$\,$yr$^{-1}$]. \\
               \texttt{pmdec} &                                                      -86.971 &                  Gaia DR2 proper motion $\mu_\delta$ [mas$\,$yr$^{-1}$]. \\
  \texttt{phot\_g\_mean\_mag} &                                                         6.85 &                                                  Gaia DR2 $G$ magnitude. \\
 \texttt{phot\_bp\_mean\_mag} &                                                        6.409 &                                      Gaia DR2 $G_\mathrm{BP}$ magnitude. \\
 \texttt{phot\_rp\_mean\_mag} &                                                        7.189 &                                      Gaia DR2 $G_\mathrm{RP}$ magnitude. \\
             \texttt{cluster} &                  NASAExoArchive\_ps\_20210506,Uma,IR\_excess &                                   Comma-separated cluster or group name. \\
                 \texttt{age} &                                                 9.48,nan,nan &  Comma-separated logarithm (base-10) of reported$^{\rm a}$ age in years. \\
           \texttt{mean\_age} &                                                         9.48 &                           Mean (ignoring NaNs) of $\texttt{age}$ column. \\
       \texttt{reference\_id} &       NASAExoArchive\_ps\_20210506,Ujjwal2020,CottenSong2016 &                          Comma-separated provenance of group membership. \\
  \texttt{reference\_bibcode} &  2013PASP..125..989A,2020AJ....159..166U,2016ApJS..225...15C &                   ADS bibcode corresponding to $\texttt{reference\_id}$. \\
\enddata


\tablecomments{
Table~\ref{tab:v06} is published in its entirety in a machine-readable
format.   This table is a concatenation of the studies listed in
Table~\ref{tab:metadata}.  One entry is shown for guidance regarding
form and content.  In this particular example, the star has a cold
Jupiter on a 16 year orbit, HD 150706b \citep{2012AA...545A..55B}.  An
infrared excess has been reported \citep{CottenSong2016}, and the star
was identified by \citet{Ujjwal2020} as a candidate UMa moving group
member ($\approx 400\,{\rm Myr}$; \citealt{mann_tess_2020}).  The
star's RV activity and TESS rotation period corroborate its youth.
}
\vspace{-0.5cm}
\end{deluxetable*}

%% file: t2_appendix.tex
\begin{deluxetable*}{lccc}
    

\tabletypesize{\scriptsize}


\tablecaption{Provenances of Young and Age-dateable Stars.}
\label{tab:metadata}


\tablehead{
  \colhead{Reference} &
  \colhead{$N_{\rm Gaia}$} &
  \colhead{$N_{\rm Age}$} &
  \colhead{$N_{G_{\rm RP}<16}$}
}

%
\startdata
                           \citet{Kounkel2020}  &             987376 &            987376 &                775363 \\
                     \citet{CantatGaudin2020a}  &             433669 &            412671 &                269566 \\
                     \citet{CantatGaudin2018a}  &             399654 &            381837 &                246067 \\
                      \citet{KounkelCovey2019}  &             288370 &            288370 &                229506 \\
                     \citet{CantatGaudin2020b}  &             233369 &            227370 &                183974 \\
                           \citet{Zari2018} UMS &              86102 &                 0 &                 86102 \\
                  \citet{SIMBAD} $\texttt{Y*?}$ &              61432 &                 0 &                 45076 \\
                           \citet{Zari2018} PMS &              43719 &                 0 &                 38435 \\
\citet{GaiaCollaboration2018} $d>250\,{\rm pc}$ &              35506 &             31182 &                 18830 \\
                      \citet{CastroGinard2020}  &              33635 &             24834 &                 31662 \\
                              \citet{Kerr2021}  &              30518 &             25324 &                 27307 \\
                  \citet{SIMBAD} $\texttt{Y*O}$ &              28406 &                 0 &                 16205 \\
                        \citet{VillaVelez2018}  &              14459 &             14459 &                 13866 \\
                     \citet{CantatGaudin2019a}  &              11843 &             11843 &                  9246 \\
                        \citet{Damiani2019} PMS &              10839 &             10839 &                  9901 \\
                                \citet{Oh2017}  &              10379 &                 0 &                 10370 \\
                          \citet{Meingast2021}  &               7925 &              7925 &                  5878 \\
                 \citet{SIMBAD} $\texttt{pMS*}$ &               5901 &                 0 &                  3006 \\
\citet{GaiaCollaboration2018} $d<250\,{\rm pc}$ &               5378 &               817 &                  3968 \\
                           \citet{Kounkel2018}  &               5207 &              3740 &                  5207 \\
                        \citet{Ratzenbock2020}  &               4269 &              4269 &                  2662 \\
                  \citet{SIMBAD} $\texttt{TT*}$ &               4022 &                 0 &                  3344 \\
                        \citet{Damiani2019} UMS &               3598 &              3598 &                  3598 \\
                           \citet{Rizzuto2017}  &               3294 &              3294 &                  2757 \\
            \citet{NASAExoArchive_ps_20210506}  &               3107 &               868 &                  3098 \\
                              \citet{Tian2020}  &               1989 &              1989 &                  1394 \\
                           \citet{Goldman2018}  &               1844 &              1844 &                  1783 \\
                        \citet{CottenSong2016}  &               1695 &                 0 &                  1693 \\
                            \citet{Gagne2018a}  &               1429 &                 0 &                  1389 \\
             \citet{RoserSchilbach2020} Psc-Eri &               1387 &              1387 &                  1107 \\
            \citet{RoserSchilbach2020} Pleiades &               1245 &              1245 &                  1019 \\
                  \citet{SIMBAD} $\texttt{TT?}$ &               1198 &                 0 &                   853 \\
                            \citet{Gagne2018c}  &                914 &                 0 &                   913 \\
                          \citet{Pavlidou2021}  &                913 &               913 &                   504 \\
                            \citet{Gagne2018b}  &                692 &                 0 &                   692 \\
                            \citet{Ujjwal2020}  &                563 &                 0 &                   563 \\
                             \citet{Gagne2020}  &                566 &               566 &                   351 \\
                      \citet{EsplinLuhman2019}  &                377 &               443 &                   296 \\
                     \citet{Roccatagliata2020}  &                283 &               283 &                   232 \\
                          \citet{Meingast2019}  &                238 &               238 &                   238 \\
                 \citet{Furnkranz2019} Coma-Ber &                214 &               214 &                   213 \\
           \citet{Furnkranz2019} Neighbor Group &                177 &               177 &                   167 \\
                             \citet{Kraus2014}  &                145 &               145 &                   145 \\
\enddata


\tablecomments{
Table~\ref{tab:metadata} describes the provenances for the young and
age-dateable stars in Table~\ref{tab:v06}.  $N_{\rm Gaia}$: number of
Gaia stars we parsed from the literature source.  $N_{\rm Age}$:
number of stars in the literature source with ages reported.
$N_{G_{\rm RP}<16}$: number of Gaia stars we parsed from the
literature source with either $G_{\rm RP}<16$, or a parallax S/N
exceeding 5 and a distance closer than 100\,pc.  The latter criterion
included a few hundred white dwarfs that would have otherwise been
neglected.  Some studies are listed multiple times if they contain
multiple tables.  \citet{SIMBAD} refers to the \texttt{SIMBAD}
database.
}
\vspace{-0.5cm}
\end{deluxetable*}

%% file: phot_table.tex
\startlongtable
\begin{deluxetable*}{cccc}
    

\tabletypesize{\scriptsize}


\tablecaption{MuSCAT3 photometry of \sn.}
\label{tab:phot}


\tablehead{
  \colhead{Time [BJD$_\mathrm{TDB}$]} &
  \colhead{Rel{.} Flux} &
  \colhead{Rel{.} Flux Err{.}} &
  \colhead{Bandpass}
}

\startdata
  2459433.829202 & 0.99719 & 0.00091 & g \\
  2459433.829324 & 0.99849 & 0.00112 & r \\
  2459433.829117 & 0.99611 & 0.00116 & i  \\
  2459433.829406 & 0.99941 & 0.00136 & z \\
\enddata


\tablecomments{
Table~\ref{tab:phot} is published in its entirety in a
machine-readable format.  Example entries are shown for guidance
regarding form and content.
}
\vspace{-0.5cm}
\end{deluxetable*}

%% file: rv_table.tex
\startlongtable
\begin{deluxetable*}{llll}
    

\tabletypesize{\scriptsize}


\tablecaption{\sn\ radial velocities.}
\label{tab:rv}


\tablehead{
  \colhead{Time [BJD$_\mathrm{TDB}$]} &
  \colhead{RV [m$\,$s$^{-1}$]} &
  \colhead{$\sigma_{\rm RV}$ [m$\,$s$^{-1}$]} & 
  \colhead{S-value}
}

\startdata
 2459433.801306 & 152.97 & 12.29 & 0.7355 \\
 2459433.812255 & 100.5 & 13.23 & 0.7434 \\
\enddata


\tablecomments{
Table~\ref{tab:rv} is published in its entirety in a
machine-readable format. 
Example entries are shown for guidance regarding form and content.
}
\end{deluxetable*}